\documentclass[a4paper,twocolumn,preprint,sort&compress]{elsarticle}

\usepackage{graphicx}
\usepackage{dcolumn}
\usepackage{bm}
\usepackage{amsmath}
\usepackage{amssymb}
\usepackage{natbib}
\usepackage{tabularx}

\begin{document}

\newcommand{\gm}{\ensuremath{g-2}\xspace}

\twocolumn[{\begin{frontmatter}
	\rightline{FERMILAB-PUB-21-069-AD-E}
	\rightline{\it{accepted by Nucl. Inst. and Meth. in Phys. Res. Sec. A}}
		
	\title{The fast non-ferric kicker system for the Muon $g-2$ Experiment at Fermilab}

	\author[1,2]{A.P.~Schreckenberger$^{1,}$}
	\cortext[cor1]{Corresponding author. Email contact:}
	\ead{wingmc@fnal.gov}
	\address[1]{Department of Physics, Boston University, Boston, MA 02215, USA}
	\address[2]{Department of Physics, University of Illinois, Urbana, IL 61801, USA}

	\author[3]{D.~Allspach}
	\author[3]{D.~Barak}
	\author[3]{J.~Bohn}
	\author[3]{C.~Bradford}
	\address[3]{
	Fermi National Accelerator Laboratory, Batavia, IL 60510, USA
	}

	\author[4,5]{D.~Cauz}
	\address[4]{INFN, Sezione di Trieste e G.C. di Udine, Trieste, Italy}
	\address[5]{Universit{\`a} di Udine, Udine, Italy}

	\author[6,7]{S.~P.~Chang}
	\address[6]{Department of Physics, KAIST, Daejeon 34141, Republic of Korea}
	\address[7]{Center for Axion and Precision Physics Research, IBS, Daejeon 34051, Republic of Korea}

	\author[8]{A.~Chapelain}
	\address[8]{CLASSE, Cornell University, Ithaca, NY 14853, USA}

	\author[3]{S.~Chappa}
	\author[3]{S.~Charity}

	\author[9]{R.~Chislett}
	\address[9]{Department of Physics and Astronomy, University College London, London WC1E 6BT, UK}

	\author[3]{J.~Esquivel}

	\author[10,11]{C.~Ferrari}
	\author[10,11]{A.~Fioretti}
	\author[10,11]{C.~Gabbanini}
	\address[10]{INFN, Sezione di Pisa, Pisa, Italy}
	\address[11]{Istituto Nazionale di Ottica, CNR-INO, S.S. A. Gozzini, via Moruzzi 1, 56124 Pisa, Italy}

	\author[10]{M.~D.~Galati}

	\author[8]{L.~Gibbons}

	\author[12]{J.~L.~Holzbauer}
	\address[12]{Department of Physics and Astronomy, University of Mississippi, University, MS 38677, USA}

	\author[10]{M.~Incagli}

	\author[3]{C.~Jensen}

	\author[13]{J.~Kaspar}
	\address[13]{CENPA, University of Washington, Seattle, WA 98195, USA}

	\author[14]{D.~Kawall}
	\address[14]{Department of Physics, University of Massachusetts, Amherst, MA 01003, USA}

	\author[15]{A.~Keshavarzi}
	\address[15]{Department of Physics and Astronomy, University of Manchester, Manchester M13 9PL, UK}

	\author[14]{D.~S.~Kessler}

	\author[3]{B.~Kiburg}
	\author[3]{G.~Krafczyk}
	\author[3]{R.~Madrak}

	\author[8]{A.~A.~Mikhailichenko$^{2,}$}

	\author[3]{H.~Nguyen}
	\author[3]{K.~Overhage}

	\author[7]{S.~Park}

	\author[3]{H.~Pfeffer}
	\author[3]{C.~C.~Polly}
	\author[3]{M.~Popovic}
	\author[3]{R.~Rivera}

	\author[1]{B.~L.~Roberts}

	\author[8]{D.~Rubin}

	\author[7,6]{Y.~K.~Semertzidis}

	\author[3]{J.~Stapleton}
	\author[3]{C.~Stoughton}
	\author[3]{E.~Voirin}
	\author[3]{D.~Wolff}

\date{\today}

\begin{abstract}
We describe the installation, commissioning, and characterization of the new injection kicker system in the Muon $g-2$ Experiment (E989) at Fermilab, which makes a precision measurement of the muon magnetic anomaly. Three Blumlein pulsers drive each of the 1.27-m-long non-ferric kicker magnets, which reside in a storage ring vacuum (SRV) that is subjected to a 1.45 T magnetic field. The new system has been redesigned relative to Muon $g-2$'s predecessor experiment, and we present those details in this manuscript.
\end{abstract}

\end{frontmatter}
}]

\footnotetext[1]{Corresponding author. Email: wingmc@fnal.gov}
\footnotetext[2]{Deceased.}

\section{\label{sec:level1}Overview}
Based at Fermilab (FNAL), the Muon $g-2$ Experiment (E989) will measure the muon magnetic anomaly ($a_\mu$) to high precision~\cite{tdr}. E989 is the successor experiment to Brookhaven National Laboratory (BNL) E821, which measured $a_\mu$ to a precision of 540 parts-per-billion (ppb) using the same 1.45 T storage ring magnet~\cite{danby}. The BNL experiment is described in Refs.~\cite{e821-1,e821-2,e821-3} where an apparent discrepancy between the theoretically calculated and experimentally measured values of the magnetic anomaly was observed. To clarify this discrepancy between the Standard Model prediction and the BNL E821 result, the FNAL experiment will make a more precise measurement of $a_{\mu}$ using a larger data set and improved systematic treatment. The first result from the Muon $g-2$ Experiment is presented in detail in Refs.~\cite{prl,omega,field,bds}. This paper describes the design, operation, and characterization of the new non-ferric fast kicker magnets that deflect the incident muon beam onto storable orbits within the ring.
\begin{figure}[!h]
	\centering
	\includegraphics[width=\columnwidth]{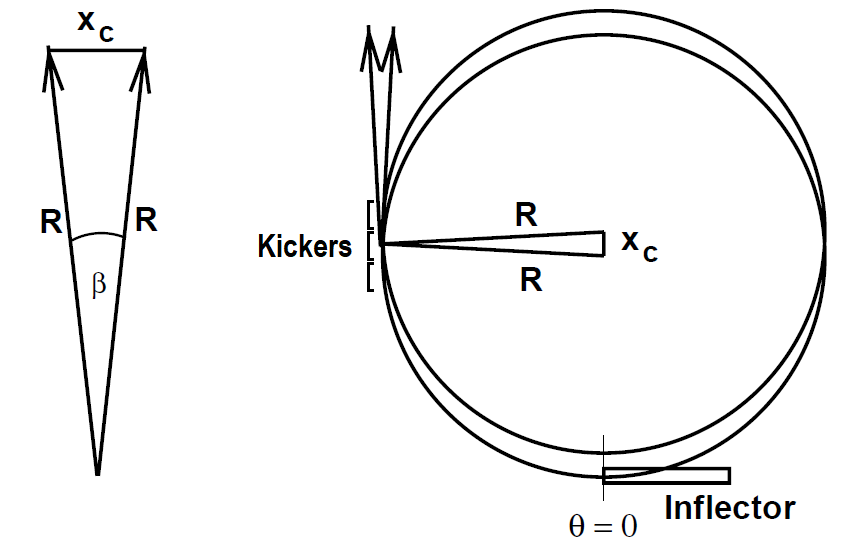}
	\caption{An idealized sketch of the beam geometry before and after the kicker pulse. $x_c$ is the 77 mm displacement of the inflector aperture relative to the central radius of the muon storage region, R=7112 mm. $\beta\approx$ 10.8 mrad. Modified from~\cite{og}. \label{fig:geometry}}
\end{figure}

The Fermilab accelerator complex produces a polarized muon beam for the experiment's consumption by colliding protons with a nickel-based target. Pions are created in these collisions that eventually decay to produce the $\mu^+$ particles that the experiment uses for physics data collection. The muons enter the storage ring vacuum (SRV) through a superconducting inflector magnet~\cite{yamamoto} that is aligned to the tangent of the ring. The inflector's interior aperture is displaced 77 mm from the central radius of the storage region. Consequently, the injected muons cannot be stored without an appropriate kick.

A series of three 1.27-m-long non-ferric aluminum kicker electromagnets are placed a quarter of the betatron wavelength from the inflector to reduce the impact of the ring magnetic field when they are pulsed with $3-4$ kA. The result of the localized perturbation moves muons onto stable orbits that facilitate a measurement of $a_\mu$. Likewise, the use of non-ferric materials was explicitly selected to not affect the $a_\mu$ measurement, which is sensitive to magnetic field sources. Beam arrives in a bunch train of 120-ns pulses at an instantaneous rate of 100 Hz (11.4 Hz average), and each bunch propagates around the ring every 149.2 ns. The kicker current must support the full duration of each beam pulse and ideally ceases before the muons make one full revolution of the ring. To address these timing specifications, Blumlein pulse forming networks (PFNs) were selected to drive the kicker currents at FNAL. This pulsing system and the kicker plate geometry provide fundamental differences from the BNL E821 infrastructure. 

Figure~\ref{fig:geometry} summarizes the effect of the kicker magnets on the beam geometry. Muons enter the ring through the inflector at $\theta = 0$ on a trajectory that would not facilitate beam storage. At around $\theta = \pi/2$, muons cross the central storage radius at an angle $\beta\approx$ 10.8 mrad. With a roughly 1.1 kG-m integrated field perturbation, the kicker magnets compensate for the $x_c$ displacement and beam is stored. The following sections will expand upon this simple picture by describing the various mechanical and electrical components of the kicker system. In addition, we describe upgrades where appropriate to emphasize the differences between the Run-1 ({\it Jan.~to~Jul. 2018}) and Run-2 ({\it started Mar. 2019}) assemblies. The system will also be characterized, and both the impact on muon storage and the kicker magnetic field will be quantified. The concluding studies on beam dynamics shown in Subsection 3.4 stress the importance of the upgrade and characterization efforts described in this manuscript. We demonstrate the relationship between kicker system performance, muon storage, and stored-beam quality, all of which motivated the design choices made by the experiment.

\section{\label{sec:level1}Kicker Design}
\begin{figure*}[!h]
	\centering
	\includegraphics[width=2\columnwidth]{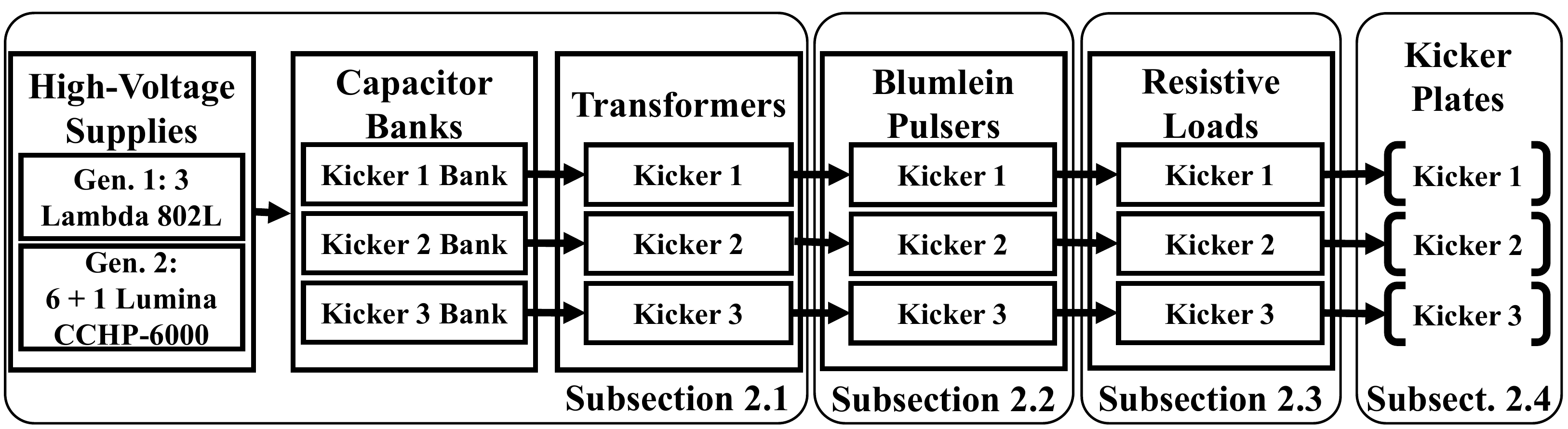}
	\caption{Flow diagram of the primary kicker subsystems. From left to right, the diagram outlines the generation of the kicker pulses from their origins in the charging supplies to when current traverses the kicker plates. The relevant subsections for each component are labeled in the curved rectangular boxes.\label{fig:BD}}
\end{figure*}
In addition to the three Blumlein PFNs, the kicker magnets rely upon a set of peripheral subsystems to generate the magnetic field perturbation needed to store muon bunches in the SRV. These subsystems include: {\it HV power supplies}, {\it capacitor banks}, {\it transformers}, {\it silicon oil supplies}, {\it castor oil pumps}, {\it resistive loads}, {\it Fluorinert chillers}, {\it low-conductivity water (LCW) connections},  {\it SRV feedthroughs},  {\it controls}, and {\it trigger timing modules}. The following four subsections will mechanically and electrically detail the major elements of the FNAL kicker. The order of the subsections is arranged to follow the kicker pulse chronologically from the initial charge until the current traverses the plates in the SRV. Figure~\ref{fig:BD} shows a flow diagram of the main kicker subsystems following that same chronology from left to right, and the associated subsections detailing each component are labeled.

\subsection{\label{sec:level2}HV Supplies, Capacitor Banks, and Transformers}

A suite of HV supplies, capacitor banks, and three transformers creates the backbone of the kicker charging circuit. Prior to the start of Run-2, three Lambda 802L supplies initiated the charging of the kicker system. In the current configuration, seven Lumina Power Inc. CCHP-6000 power supplies, rated for 1 kV (6 kW) operations, were installed to increase the peak charging current over the 802L units. Each of the kicker magnets has an associated capacitor bank, which receives charge from two Lumina supplies connected in parallel. The seventh supply serves as a hot spare. Upon receiving the charging signal from the controls rack, the HV supplies deliver current to the capacitor banks for a period of approximately 9 ms.

The three capacitor banks consist of five 50 $\mu$F Genteq 97F5211S capacitors. The elements are connected in parallel to produce 250 $\mu$F of total capacitance, though the option to expand up to 500 $\mu$F is available. After the banks are charged, another control signal triggers a silicon controlled rectifier (SCR) to permit the flow of current from the selected bank to the allocated single-phase transformer. 

Upon design for Run-1, the expected gains of the transformers were 1:85. Each transformer is submerged in silicon oil and housed in an aluminum vessel. In each vessel, we mounted a capacitive divider that reads out the voltage from the secondary winding. It is from these monitors that we obtain the voltages placed on the PFNs, which are critical measurements for calibrating the amplitude of the kicker pulses.  

Beamline studies conducted during Run-1 repeatedly indicated that the kickers were operating at a lower voltage than the initial calibration suggested. An audit of the system revealed two transformer system characteristics that differed from specifications: 1) the gains of the transformers were not 1:85, and 2) the capacitive dividers had gains that differed by as much as 43.3$\%$ from the expected ratio of 5000:1.

The final calibrations presented in Subsection 3.1 mitigate the variances due to divider behavior and cable capacitance by mapping the responses of the capacitive dividers in the transformer vessels to the response of a calibrated resistive divider. Additional testing performed just prior to the start of Run-2 verified the earlier findings. We summarize the results of these transformer tests in Table~\ref{tab:trans}.
\begin{table*}[!h]
	\centering
	\caption{\label{tab:trans}Voltage and inductance measurements of the K2 and K3 Blumlein charging transformers. These observations verified the calibration effort that is described in Subsection 3.1 by directly obtaining the gain ratios of the transformers and their capacitive divider monitors.}
	
	\begin{tabular}{llll}
		\hline
		\hline
		Diagnostic &K2 Test 1 &K2 Test 2 &K3 Test 1\\
		\hline
		Leakage inductance &183$\mu$H &180$\mu$H &230$\mu$H\\
		Primary inductance &100mH &100mH &100mH\\
		\hline
		Input voltage &11.3V$_{pp}$ & 23.4V$_{pp}$ & 22.2V$_{pp}$\\
		Output voltage &0.99kV$_{pp}$ &2.12kV$_{pp}$ &2.08kV$_{pp}$\\
		Ratio &1:87.8 &1:90.6 &1:93.7\\
		\hline
		Secondary input voltage &0.99kV$_{pp}$ &2.20kV$_{pp}$ &2.02kV$_{pp}$\\
		Secondary output voltage &0.35V$_{pp}$ &0.74V$_{pp}$& 0.58V$_{pp}$\\
		Capacitive divider ratio &2834.3:1 & 2973.0:1 & 3458.9:1\\
		\hline
		\hline
	\end{tabular}
	
\end{table*}

\subsection{\label{sec:level2}Blumlein Pulsers}

\begin{figure}[!h]
\includegraphics[width=\columnwidth]{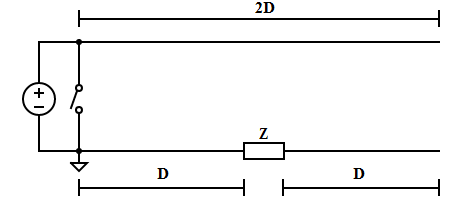}
\caption{The simplest circuit diagram for a Blumlein pulser using series coaxial lines. The load, Z, is attached on the ground conductor at the midpoint of an unterminated line with length 2D.\label{fig:blumsimple}}
\end{figure}

Triaxial Blumlein pulsers were introduced in the Muon $g-2$ Experiment as a major improvement over the underdamped RLC PFN utilized in BNL E821. Invented in 1937 by Alan Blumlein, Blumlein pulsers provide several advantages when compared to a standard transmission line~\cite{blum,blum-2}. As demonstrated by the illustration of a Blumlein pulser in Fig.~\ref{fig:blumsimple}, the switching element in this type of PFN can be grounded, which simplifies triggering electronics and offers an additional layer of safety. Another advantage of the Blumlein PFN is that this system yields a pulse with a peak voltage equal to that of the power supply output. A standard transmission line would require double the power supply voltage to produce the same voltage drop across the load. Furthermore, with a properly matched load impedance, reflections from the unterminated conductors are exploited in Blumleins to promptly end the electrical pulse. In BNL E821, the RLC PFN generated a primary pulse that exceeded 450 ns and engulfed the cyclotron structure of the muon beam~\cite{tdr,og}. We selected the Blumlein PFN design specifically to improve the timing characteristics of the kicker pulse, the duration of which is shown in Eq.~\ref{eq:time} (where $L$ is the length of the Blumlein). 
\begin{equation}
\tau = 2L\frac{\sqrt{\mu_r\varepsilon_r}}{c}.
\label{eq:time}
\end{equation}

\begin{figure*}[]
	\centering
	\includegraphics[width=1.9\columnwidth]{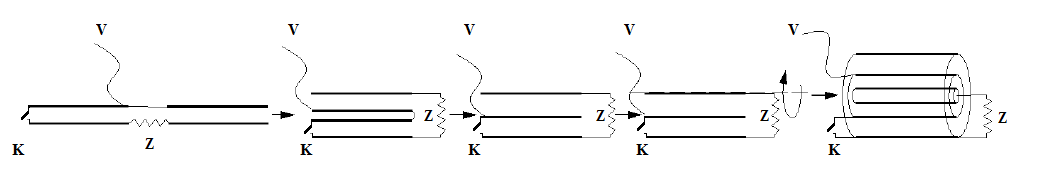}
	\caption{Topological modification of the series coaxial Blumlein model shown in Fig.~\ref{fig:blumsimple} to the triaxial configuration used at Fermilab. V represents the charging voltage of the Blumlein. K is the switching element, and Z is the load impedance.\label{fig:topblum}}
	
	\includegraphics[width=1.9\columnwidth]{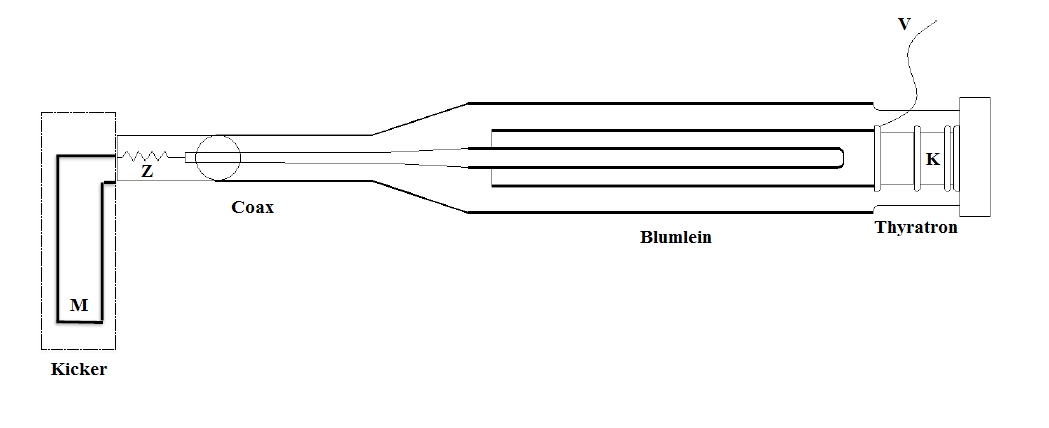}
	\caption{Representation of the triaxial Blumlein assembly with the kicker magnet, M, and resistive load, Z. The load impedance is connected to the small and large Blumlein conductors through a coaxial cable. The medium Blumlein conductor is charged to voltage V. The switching element, K, is a thyratron that facilitates the short between the medium and large conductors. \label{fig:blumcom}}
	
	\includegraphics[width=1.9\columnwidth]{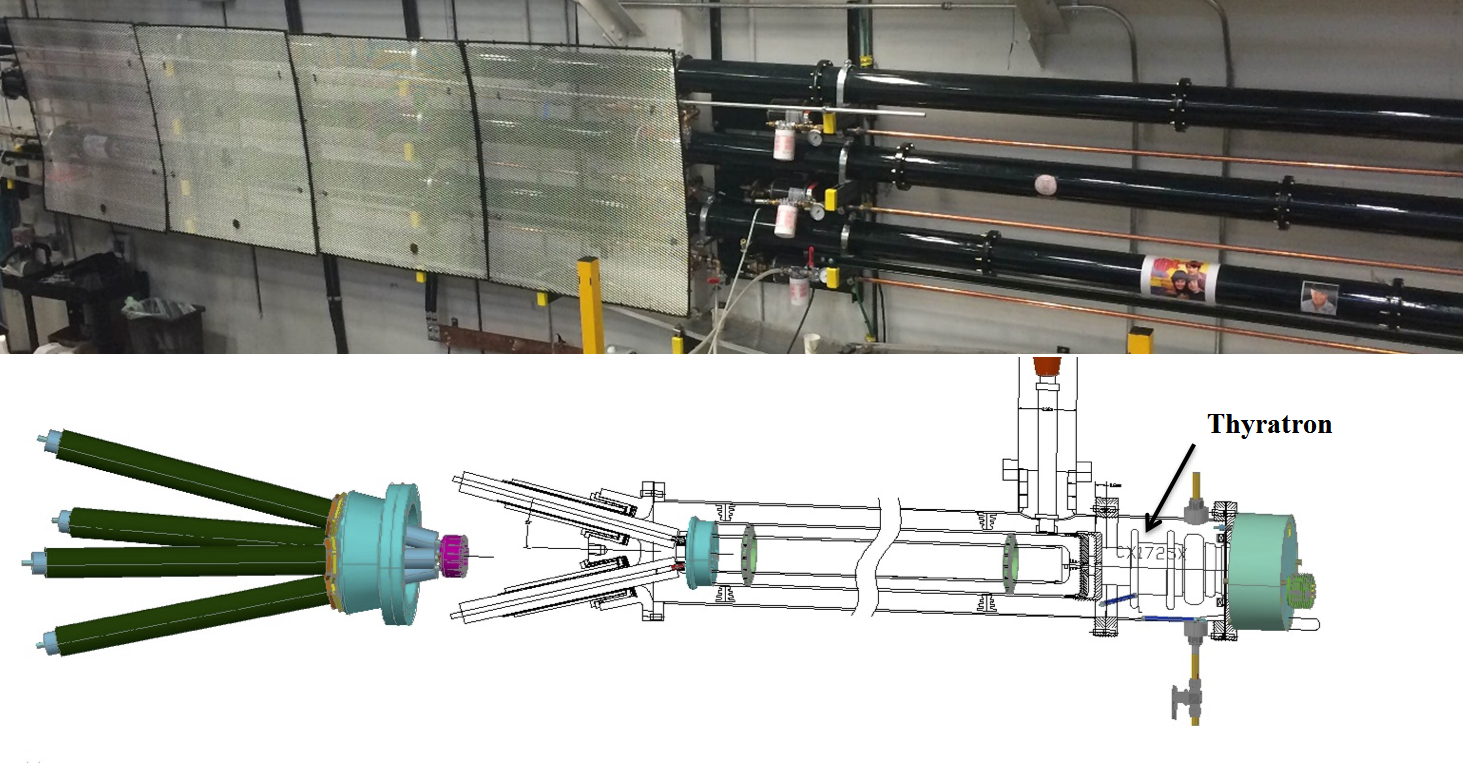}
	\caption{{\it Top:} Picture of all three 9.35-m-long Blumlein PFNs installed in the Muon $g-2$ experimental hall. {\it Bottom:} Illustration of the Blumlein assembly with all three conductors. The attached CX1725X thyratron is visible, and the connection of the 5 k$\Omega$ resistor is depicted immediately to the left of the thyratron. At the left of the diagram, a prototype output cable assembly is drawn. A modified design with parallel cable sockets, capable of seating up to four coaxial cables, is currently used.\label{fig:blumdraw}}
\end{figure*}

The triaxial Blumelin configuration used by the experiment can be represented through a topological modification of the circuit shown in Fig.~\ref{fig:blumsimple}. The process is illustrated in Fig.~\ref{fig:topblum}, and an expanded detailing of the Blumlein assembly is provided in Fig.~\ref{fig:blumcom}. Each 9.35-m-long Blumlein consists of three concentric cylindrical aluminum conductors ({\it small, medium, large}) that are approximately 3 mm thick. The characteristic impedance of the Blumlein ($Z_{0}^{B}$) is set by the geometry of the conductors, specifically the conductor radii, and the relative permittivity ($\varepsilon_{r}$) of any dielectric medium used in the system. The Muon $g-2$ Blumleins are filled with castor oil, which has a measured relative permittivity of 4.57. The castor oil also serves as a cooling medium for the PFN switches. The fluid is circulated through the Blumleins by pumps connected to each pulser with copper piping.

Equation~\ref{eq:z0b} is the expression for the characteristic impedance of the Blumleins. 
\begin{equation}
Z_0^B = \frac{60\Omega}{\sqrt{\varepsilon_{r}}}\ln{\frac{R}{r}},
\label{eq:z0b}
\end{equation}
where $\varepsilon_{r} = 4.57$, $R$ is the inner radius of the larger conductor (either $R= R_{large}^{inner} = 71.12 mm$ or $R = R_{medium}^{inner} = 53.72mm$), and $r$ is the outer radius of the smaller conductor (either $r = r_{medium}^{outer} = 56.77 mm$ or $r = r_{small}^{outer} = 43.18 mm$). From these possibilities, Equations~\ref{eq:z0bsm}$-$\ref{eq:z0ml} are generated, yielding the following impedance calculations. In the ideal case, the Blumlein impedance would be equal to half of the load impedance ($Z/2 \approx 6.25\Omega$).
\begin{eqnarray}
Z_0^{Bsm} &= &\frac{60\Omega}{\sqrt{4.57}}\ln{\frac{53.72}{43.18}} = 6.13\Omega,\label{eq:z0bsm}\\
Z_0^{Bml} &= &\frac{60\Omega}{\sqrt{4.57}}\ln{\frac{71.12}{56.77}} = 6.33\Omega.\label{eq:z0ml}
\end{eqnarray}

On one end of each Blumlein PFN, a thyratron is seated as the switching element between the medium and large conductors. The output cable from one transformer connects on this end of the Blumlein, as seen in Fig.~\ref{fig:blumcom}. The ground of the transformer cable attaches directly to the large conductor of the Blumlein while the high voltage core connects to the medium Blumlein conductor through a 5 k$\Omega$ resistor.

When a thyratron shorts the conductors of a Blumlein, it initiates the current pulse that drives the kicker magnet. Given the voltage and impedance characteristics of the kicker system, a peak current exceeding 4 kA is achievable. We installed E2V Technologies CX1725X thyratrons on the Blumleins. These thyratrons hold off peak anode voltages up to 70 kV. They are capable of conducting currents up to 15 kA, and their maximum allowed repetition rate is 2 kHz, which exceeds the instantaneous delivery rate of the muon beam by a factor of 20. The CX1725X model uses a tetrode layout, meaning that the triggering mechanism is controlled via two grids. One grid receives a $10-25$ A current signal to ionize the deuterium gas inside the thyratron. The other grid is held at a -0.2 kV potential to keep the thyratron in a non-conductive, inhibited state. When the discharge trigger is received from the timing system, a $+$2.0 kV pulse is delivered to the same grid as the inhibit. This moves the thyratron into its conductive state.

With all of these elements described, we can now construct the illustration in Fig.~\ref{fig:blumdraw}. Here, the diagram depicts the design of all three Blumlein conductors, the connection of the thyratron, the attachment of the input 5 k$\Omega$ resistor, and a prototype output cable assembly. Also pictured in this figure is an image of the three Blumlein PFNs mounted on the wall of the experimental hall.

\subsection{\label{sec:level2}Resistive Loads and SRV Feedthroughs}

The Blumlein PFNs deliver pulses to the kicker magnets through electrically-conducting, SRV feedthroughs. Outside the vacuum region, resistive loads electrically connected to these elements facilitate the necessary current to produce the desired kicker magnetic field.   

\begin{figure}[]
	\centering
	\includegraphics[width=0.6\columnwidth]{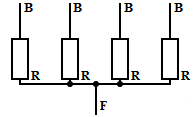}
	\caption{Schematic of the resistive load topology used in Run-1. This design consisted of independent cable-resistor connections to form a 12.5 $\Omega$ element. B represents a cable from the Blumlein PFN. Each R is a 50 $\Omega$ resistor, and F is the high voltage terminal of the vacuum feedthrough. \label{fig:bazG1}}

	\centering
	\includegraphics[width=\columnwidth]{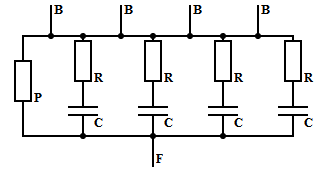}
	\caption{Schematic of the resistive load topology used in Run-2. Each Blumlein cable, B, connects to a common bus in this design. P represents eight 100 $\Omega$ resistors connected in parallel. R represents a 50 $\Omega$ resistor, and C is a 0.9 nF capacitor. The P and RC elements are installed in two concentric rings to minimize inductance. F is the high voltage terminal of the vacuum feedthrough.\label{fig:bazG2}}
\end{figure}

Two load designs were used in the experiment. The first iteration was used throughout the commissioning phase and Run-1. A schematic of the resistive load topology is included in Fig.~\ref{fig:bazG1}. In this design, four Times Microwave AA5966 coaxial cables from the Blumlein PFN were independently connected to 50 $\Omega$ resistors housed in an aluminum containment vessel. The resistors were also attached to the central, high voltage terminal of the vacuum feedthrough, which mates to the inner kicker plate through an aluminum lead. The return path follows the outer kicker plate until a lead carries the current through the load containment vessel back to the braids of the coaxial cables. The idealized result of this arrangement is the production of a 12.5 $\Omega$ load that is tuned to the impedance of the Blumlein pulser.

We developed the second iteration of the load to improve the strength of the kicker field for Run-2. Here, the AA5966 central conductors connect to a common bus that securely holds eight 100 $\Omega$ resistors in parallel with four RC elements, each consisting of a 50 $\Omega$ resistor in series with a 0.9 nF capacitor. Another bus connects the load to the high voltage input of the vacuum feedthrough. The Run-2 assemblies also possess capacitive dividers that serve as another diagnostic tool. A schematic of this load topology is included in Fig.~\ref{fig:bazG2}. 

We observe both electrical and mechanical improvements in the new load design. By using RC components, we increase the maximum amplitude of the kicker magnet current by roughly 10$\%$. Consequently, the kicker system can achieve the same peak field perturbation at a reduced Blumlein charging voltage. This effect can be understood through a basic knowledge of capacitor behavior. At some time $t_0$, the capacitors act like closed conductors, which reduces the impedance to 6.25 $\Omega$. As the capacitors reach $Q_{max}$, the load resistance asymptotically approaches 12.5 $\Omega$.
\begin{figure}[]
\includegraphics[width=\columnwidth]{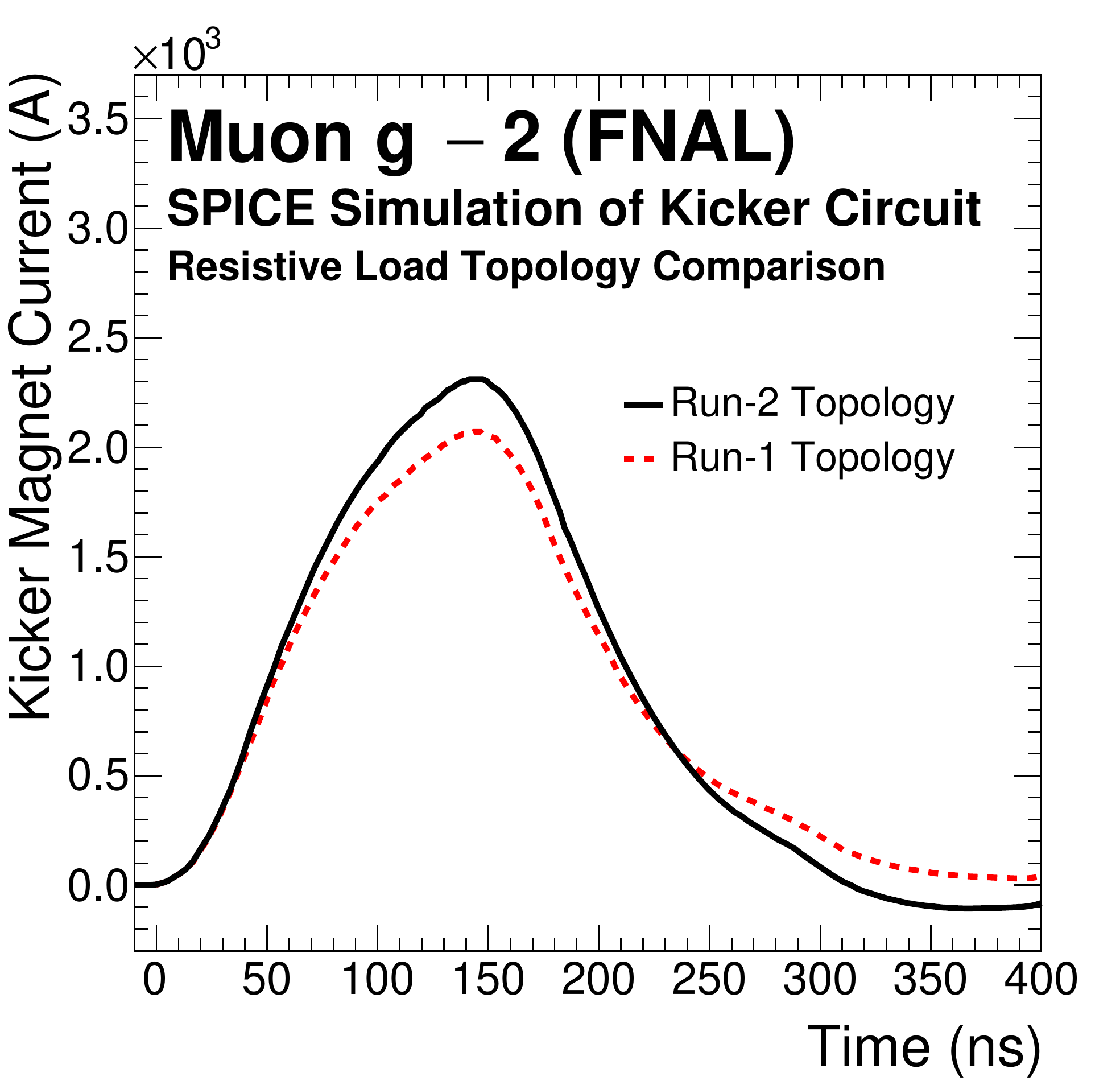}
\caption{SPICE simulation of the kicker magnet current at approximately 30 kV operation. We show current waveforms for both of the resistive load topologies. \label{fig:bazcomp}}
\end{figure}

Using SPICE~\cite{spice}, an open-source circuit simulator, we modeled the kicker system as lump elements and produced the comparison of the resistive load topologies displayed in Fig.~\ref{fig:bazcomp}. Here, we simulate that the peak amplitude of the current in the kicker magnet increases by 11.6$\%$, which agrees with our initial estimate.

\begin{figure*}[]
	\centering
\includegraphics[width=1.9\columnwidth]{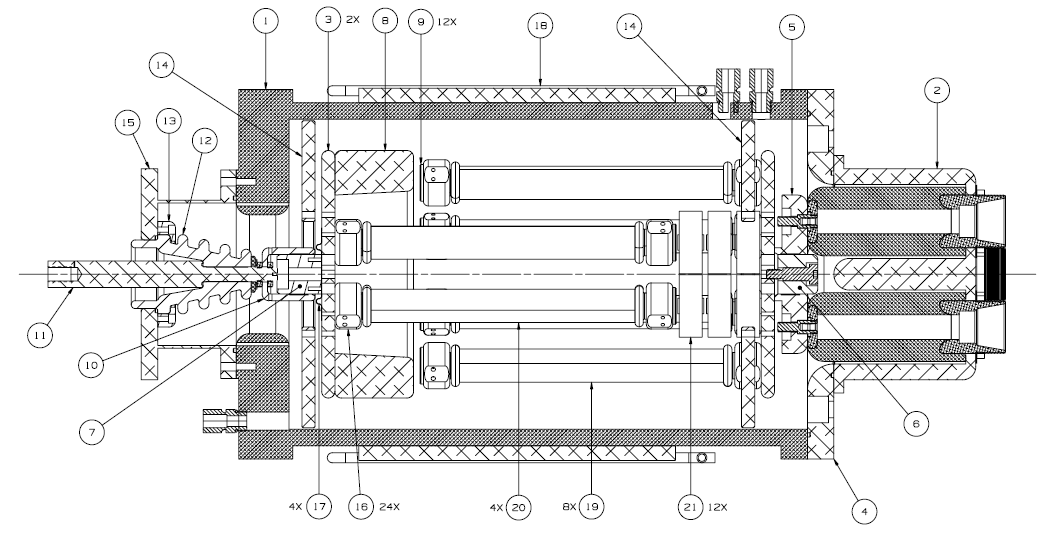}
\caption{Technical design drawing of the Run-2 load assembly~\cite{bazd}. Element 19 represents the eight 100 $\Omega$ resistors. Elements 20 and 21 comprise the RC component of the newly developed topology. The connection to the SRV feedthrough is displayed in the leftmost portion of the image (Elements 10$-$13, 15). The sockets for the input cables from the Blumlein PFN are shown in the right side of the image (Elements 2, 4$-$6).\label{fig:bazintpro}}
\end{figure*}
\begin{figure*}[]
	\centering
\includegraphics[width=1.89\columnwidth]{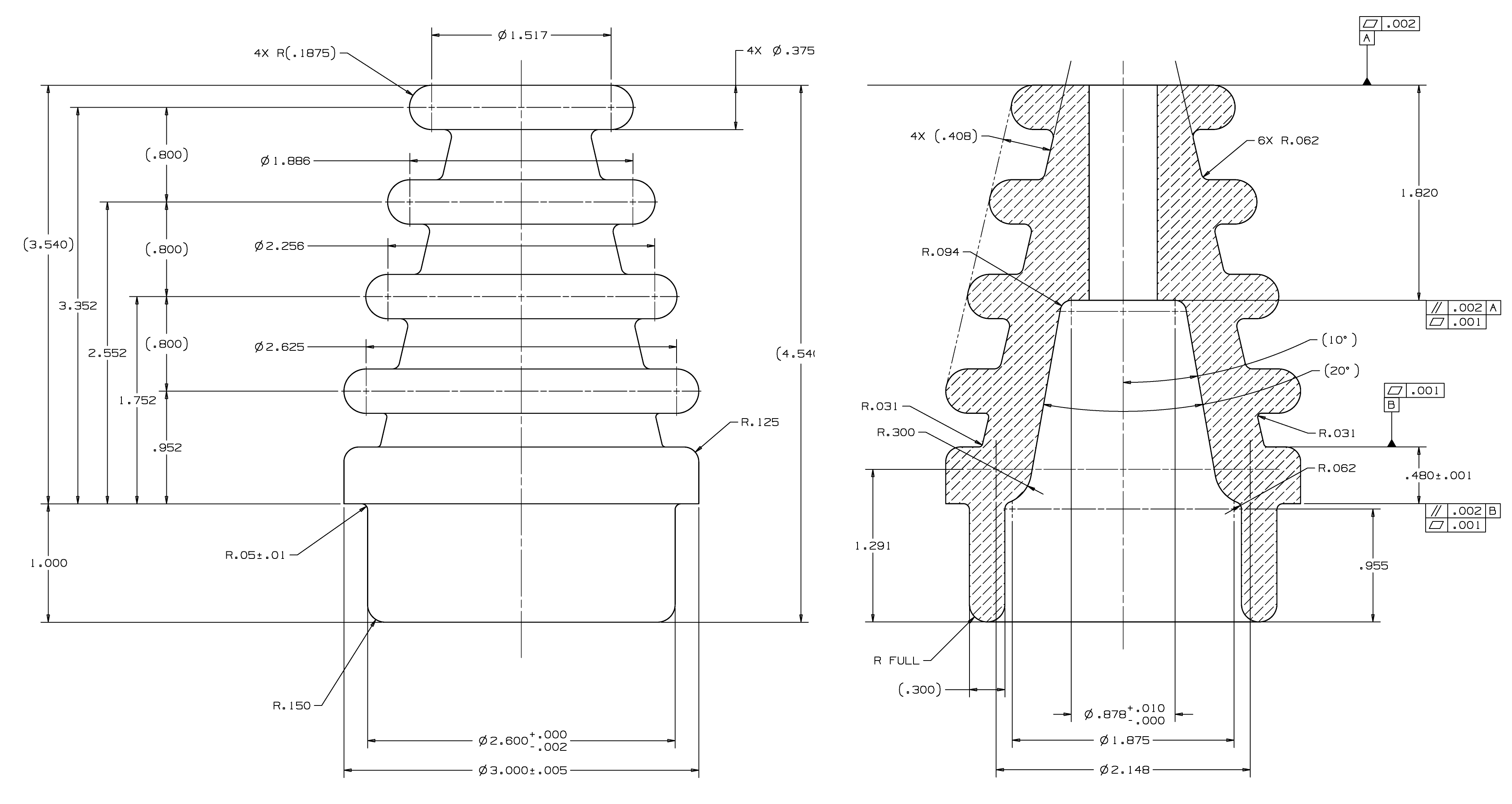}
\caption{Technical design drawing of the SRV HV feedthrough~\cite{bazf}. The HV conductor passes through a central aperture where it connects to the resistive load on the top side of the drawing and the SRV lead on the bottom side of the drawing. {\it Left:} Exterior dimensions of the Macor ceramic. {\it Right:} Interior dimensions of the Macor feedthrough. The ceramic is hatched.\label{fig:bazf}}
\end{figure*}

A detailed design drawing of the Run-2 resistive load is provided in Fig.~\ref{fig:bazintpro}. The RC components described in this section are represented by Elements 19$-$21 in the figure. In both load designs, the interiors of the housing vessels were filled with Fluorinert that served as a cooling fluid. Attached plumbing facilitated the flow of Fluorinert to chiller pumps located outside the magnetic storage ring. Considering the shape of the kicker current pulse and the accelerator duty cycle, the resistive loads will emit approximately 300 W. Each kicker assembly has a dedicated chiller that provides 200 W of cooling power. The additional heat load is dissipated through radiative blocks on the aluminum housings that are actively cooled with flowing low-conductivity water. We record measurements of the Fluorinert temperature in each of the chiller units, which yield feedback regarding the status of the loads. At nominal operation, we observe temperatures ranging from 22$^{\circ}$C$-$25$^{\circ}$C from the readout of these sensors. We installed additional failsafe mechanisms directly onto each of the load housings. If the temperature exceeds 62.8$^{\circ}$C at a resistive load, temperature-sensitive switches provide a veto trigger to prevent further pulsing.

The SRV feedthrough component of the assembly consists of a central HV conductor shrouded by a Macor ceramic insulator. The vacuum is sealed using two Viton o-rings seated between the insulator and the vacuum flange. At various times, we used indium seals in place of one of the o-rings, but the Viton-only configuration met the mechanical and electrical specifications of the experiment. We include the technical design drawings of the Macor vacuum feedthrough in Fig.~\ref{fig:bazf}. 

\subsection{\label{sec:level2}Kicker Magnets}
\begin{figure}[]
	\centering
	\includegraphics[width=\columnwidth]{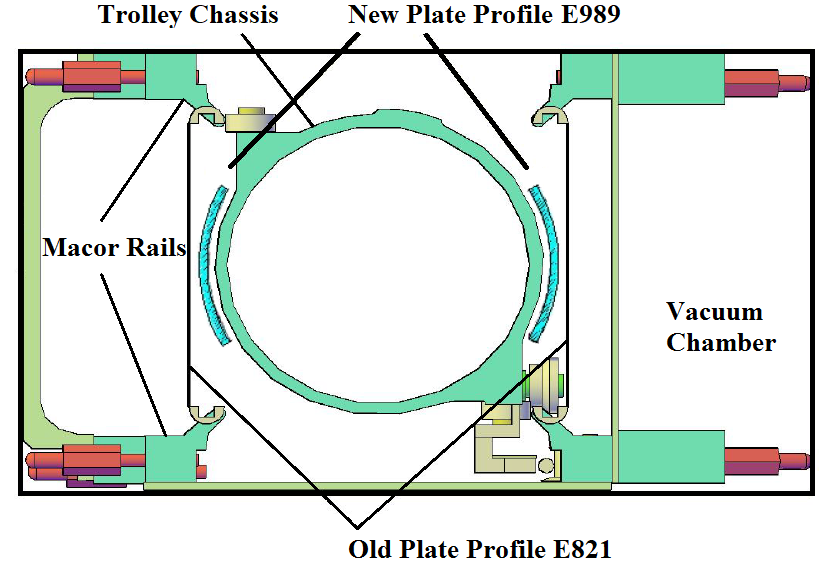}
	\caption{Elevation view of the SRV in the kicker region. We superimpose projections of the E821 kicker plates, the new kicker plates, the trolley chassis, and the Macor rails used by the trolley. The trolley requires clearance to pass by the plates to map the magnetic field in the muon storage region, and the kickers are electrically disabled when the trolley is deployed.\label{fig:pp}}
	
	\includegraphics[width=\columnwidth]{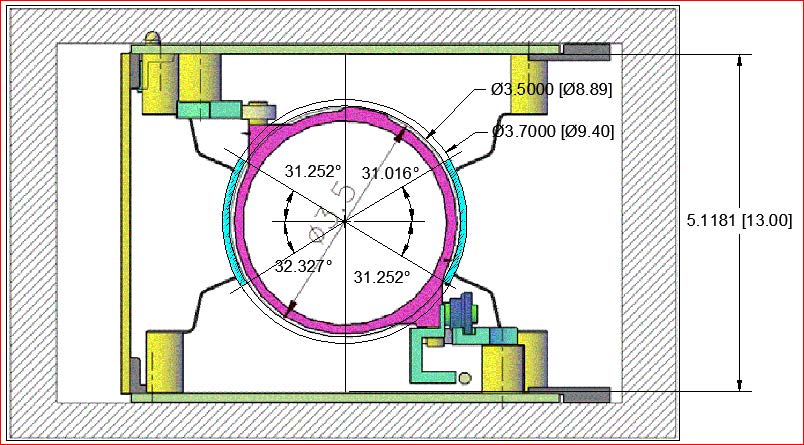}
	\caption{Elevation view of the SRV in the kicker region at FNAL. The Run-1 ceramic standoffs (tan), aluminum brackets (black), and kicker plates (cyan) are depicted along with radial and angular dimensions. The trolley rails (green) and the outline of the trolley shape (pink) are also illustrated.\label{fig:ppt}}
	
	\includegraphics[width=\columnwidth]{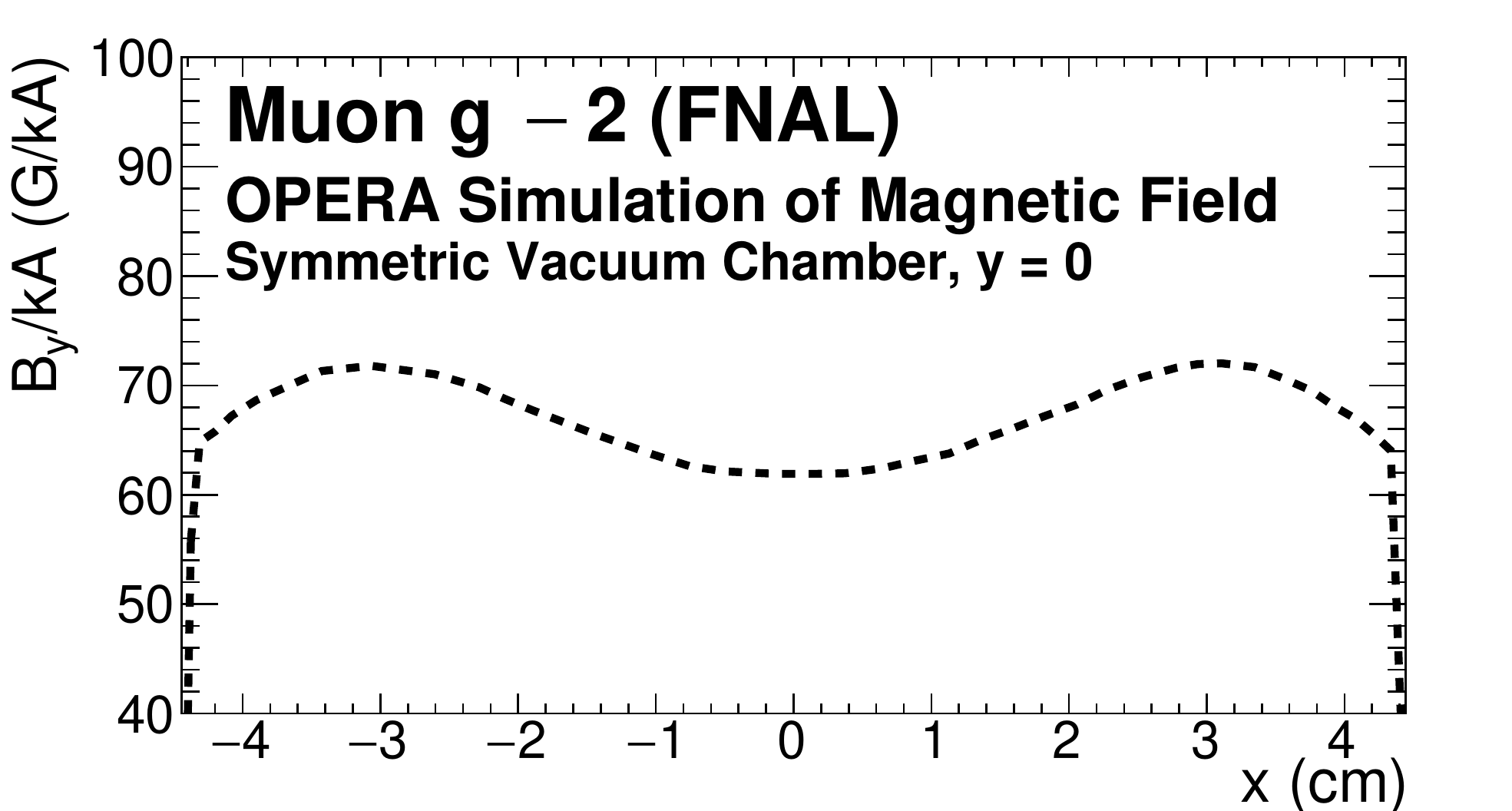}
	\caption{OPERA simulation of the kicker-induced magnetic field inside a symmetric (rectangular) vacuum chamber. We display the B$_{y}$ component of the field at y$=$0 in the beam storage region.\label{fig:opera}}
\end{figure}

\begin{figure*}[]
	\centering
	\includegraphics[width=1.9\columnwidth]{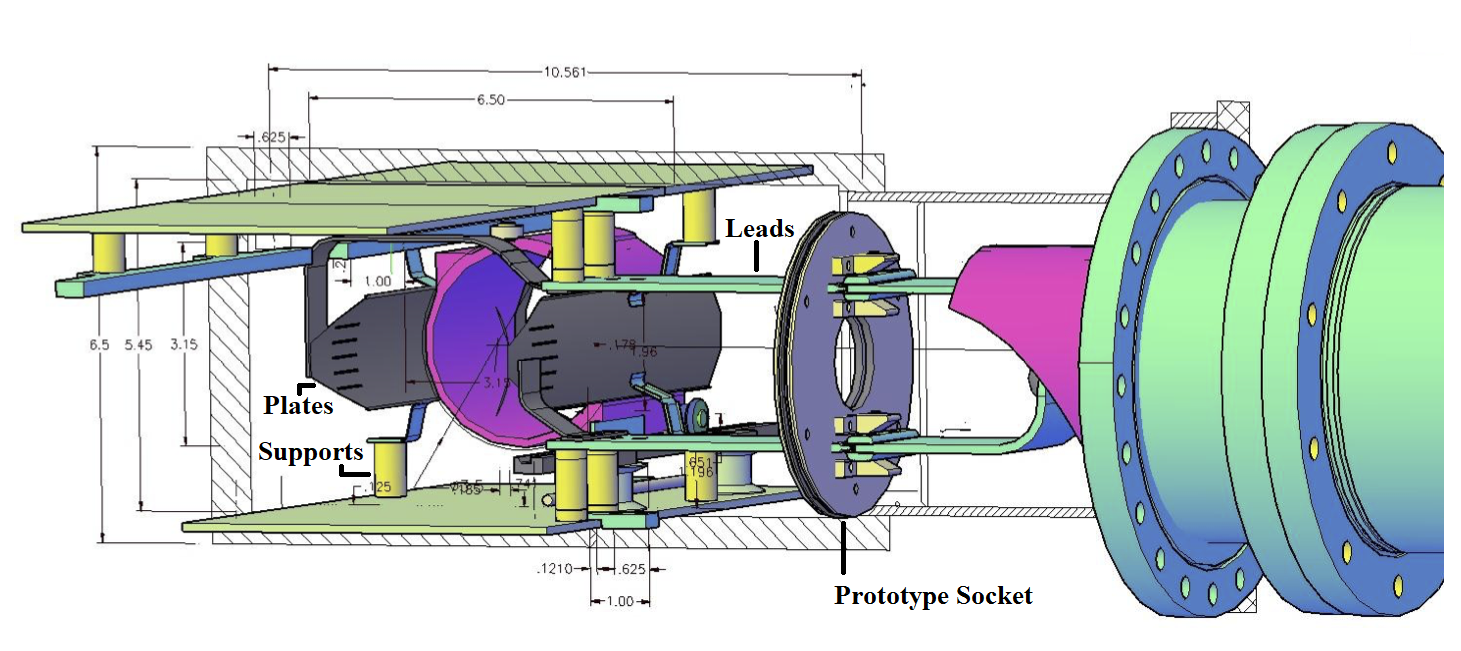}
	\caption{Isometric diagram of the SRV feedthrough, high voltage leads, and upstream end of the kicker magnet. This depiction represents the system prior to the start of Run-1. Aluminum sleds and Teflon sockets replaced the prototype socket shown. The beam storage region is displayed at left. We illustrate the kicker plates, current return, and support structures. The pictured ceramic standoffs are still used. However, we implemented a fluted design to opportunistically address breakdown points.\label{fig:itp}}
\end{figure*}

The kicker magnets each consist of a pair of curved aluminum plates that are 1.27 meters in length. Here, we exhibit another design difference between the E821 and E989 kicker systems. From the profile presented in Fig.~\ref{fig:pp}, we illustrate the changes to the kicker region of the SRV with the notable difference coming from the geometry of the Muon $g-2$ Experiment's kicker magnets. By utilizing curved plates, the new kicker system is better optimized for efficiency than the E821 counterpart. The curved profile produces a more uniform magnetic field perturbation around the center of the beam storage region. Initial simulations performed for the experiment's technical design report suggested that the modified plate geometry would produce a midplane field of 65 G/kA. This translates to an 86$\%$ improvement in the field:current efficiency when compared to the E821 plate geometry. Details regarding the design specifications of the kicker system are described in Chapter 12 of Ref.~\cite{tdr}. As part of the system characterization effort, a measurement of the kicker-induced magnetic field was performed, and the results of that analysis are included in Subsection 3.3.

To support the redesigned profile, the 0.737-mm-thick kicker plates are suspended from vacuum chamber cages using riveted aluminum brackets and ceramic standoffs. Additional riveted bridges, located at the ends of the plates, serve as the electrical connections for every pair. Several challenges arose during the implementation of the new geometry. The experiment uses a collection of mobile nuclear magnetic resonance (NMR) probes to map the magnetic field around the azimuth of the SRV. The probes are contained in a chassis, and the device is commonly referred to as the trolley. In E821, the kicker magnet assembly served as the rails for the trolley as it traversed the kicker region, but at Fermilab, a dedicated rail is required. The radius of curvature for the plates is 4.445 cm, which we determined by requiring clearance for the trolley to pass through the kicker region. The diagram in Fig.~\ref{fig:ppt} depicts only the FNAL plate geometry. In this figure, we illustrate the support framework of the kicker magnet, relevant cylindrical dimensions, and the trolley. 

Current from the Blumlein PFNs is delivered to the kicker magnets through high voltage leads that connect to the SRV feedthrough. The isometric view in Fig.~\ref{fig:itp} illustrates the kicker magnet assembly at the high voltage attachment point. This diagram shows the conditions prior to the start of Run-1. The SRV feedthrough is shown on the right side of the figure along with the connection to the leads. This design had to be modified because we determined that it was insufficient during system commissioning. Charge would continually accumulate on the insulators, and thus, the disk became a breakdown point of the system. In Run-2, we utilized aluminum sleds and smaller Teflon sockets to guide the electrical contacts of the leads. These devices have smoother surfaces and greater clearances to suppress the potential for electrical breakdowns.

The beam storage region is visible on the left side of Fig.~\ref{fig:itp}. The upstream portion of the kicker plates is drawn with the current return, several aluminum supports, and some ceramic standoffs. During Run-1, we recorded breakdowns across the ceramic standoffs that held the high voltage (innermost) plate. To address this issue, we implemented a fluted standoff design that increases the effective distance between the aluminum bracket and the SRV walls. Opportunistic installation of the modified standoffs has proven effective at increasing the kicker system uptime. As with the previous figures, the trolley rails and chassis outline are also depicted.

We performed a more recent simulation of the kicker-induced magnetic field using OPERA Version 18R2~\cite{opera}. In addition to the dimensions listed in Fig.~\ref{fig:ppt}, the model included the value of $2.5\times 10^{7}$ S/m for the conductivity of aluminum. We present the results of the simulation in Fig.~\ref{fig:opera}, where the computed midplane magnetic field is 62 G/kA. Additional studies evaluated the impact of asymmetric (scalloped) vacuum chambers, which led to a 2.4$\%$ increase in the midplane magnetic field (63.4 G/kA), and we also provide a preliminary measurement for the kicker plate inductance of 0.8 $\mu$H. This value does not include other sources of inductance in the system. We present a more thorough study of the kicker system impedances in Subsection 3.2.

\subsection{\label{sec:level2}Controls and Timing}

The controls and timing systems facilitate two fundamental interactions with the kicker. Several instruments establish links between operators, monitoring hardware, and the kicker charging circuit, whereas other elements parse information from the injection accelerator to the kicker circuit and data acquisition electronics.

To control the operating voltage of the kicker circuit, a LabJack T7 unit produces three reference voltages, $(V^{K1}_{ref},V^{K2}_{ref},V^{K3}_{ref})$, that set the output amplitudes of the HV supplies described in Subsection 2.1. A thorough calibration that maps the reference voltages to Blumlein charging voltages is described in the System Characterization section (Section 3) of this manuscript. 

Triggers for the kicker circuits are introduced via Clock and Commands Center (CCC) logic. CMS FC7 boards ({\em  FPGA-based hardware for generic data acquisition and control}) produce optical signals that are utilized by the kicker control instruments in the experimental hall. 

The first signal initiates the charging of the capacitor banks and establishes a negative bias on the thyratron grids to keep the switches in their open states. With NIM logic, the charging signal is split to produce a delayed signal for the capacitor bank SCRs, which commence the charging of the Blumlein PFNs.

Additional optical signals provide discharge triggers for the three thyratron control chassis. Upon receipt, 2 kV potentials are placed on the second thyratron grids that force the switches into their conductive states. This action drives current through the magnet load. Adjustment at the nanosecond level facilitates the precise timing of the kicker pulses relative to the injected beam. This additional flexibility aided beam storage optimization efforts. The CCC also triggers both synchronous and asynchronous data acquisition of the kicker monitoring devices, which are essential for quality assurance, data analysis, and safety.

Furthermore, both the LabJack and CCC controls are connected to custom FPGA hardware that registers anomalous electrical waveforms. Information pertaining to this specific feature of the controls hardware is described in the following subsection.  

\subsection{\label{sec:level2}Monitors}

\begin{table*}[]
	\centering
	\caption{\label{tab:mon}Monitoring systems for the Muon $g-2$ Experiment's non-ferric kicker. Source and function are also included. }
	
	\begin{tabular}{lll}
		\hline
		\hline
		Monitor &Source &Function\\
		\hline
		PFN charging voltages &Capacitive dividers &Calibration\\
		&  & Timing \\
		
		Pickup coil voltages &Coaxial loops &Spark detection\\
		&  & Pulse analysis \\
		&  & Timing \\
		
		Resistive load voltages &Capacitive dividers &Diagnostics\\
		\hline
		Hall audio & Microphones &Spark detection\\
		& & ({\it PFN, cables})\\
		
		Vacuum pressure & SRV gauges &Spark detection\\
		
		\hline
		Thyratron currents & HV supply  & Operations\\
		Castor oil flow  & PFN flow gauges & Operations\\
		Castor oil temp.  &PFN temp. gauges &Operations\\
		Fluorinert flow & Chillers &Operations\\
		Fluorinert temp. & Chillers &Operations\\
		\hline
		\hline
	\end{tabular}
	
\end{table*}

Various monitors installed on components of the kicker system collect data that serve both operational and analytical purposes. In particular, electrical waveforms pertaining to the Blumlein PFN charging voltages, load voltages, and lead currents produce information that affects analysis efforts and operation decisions. Additional readouts are directly fed into the controls rack such that triggers can be vetoed under unfavorable hardware conditions. Monitors of fluid levels, flows, and temperatures, for example, act as binary-style switches that provide input for operations. A comprehensive list of important system monitors is included in Table~\ref{tab:mon}. For the remainder of this subsection, we focus on the monitoring of critical electrical waveforms and discuss their contributions in greater detail.

\begin{figure}[!h]
	\includegraphics[width=\columnwidth]{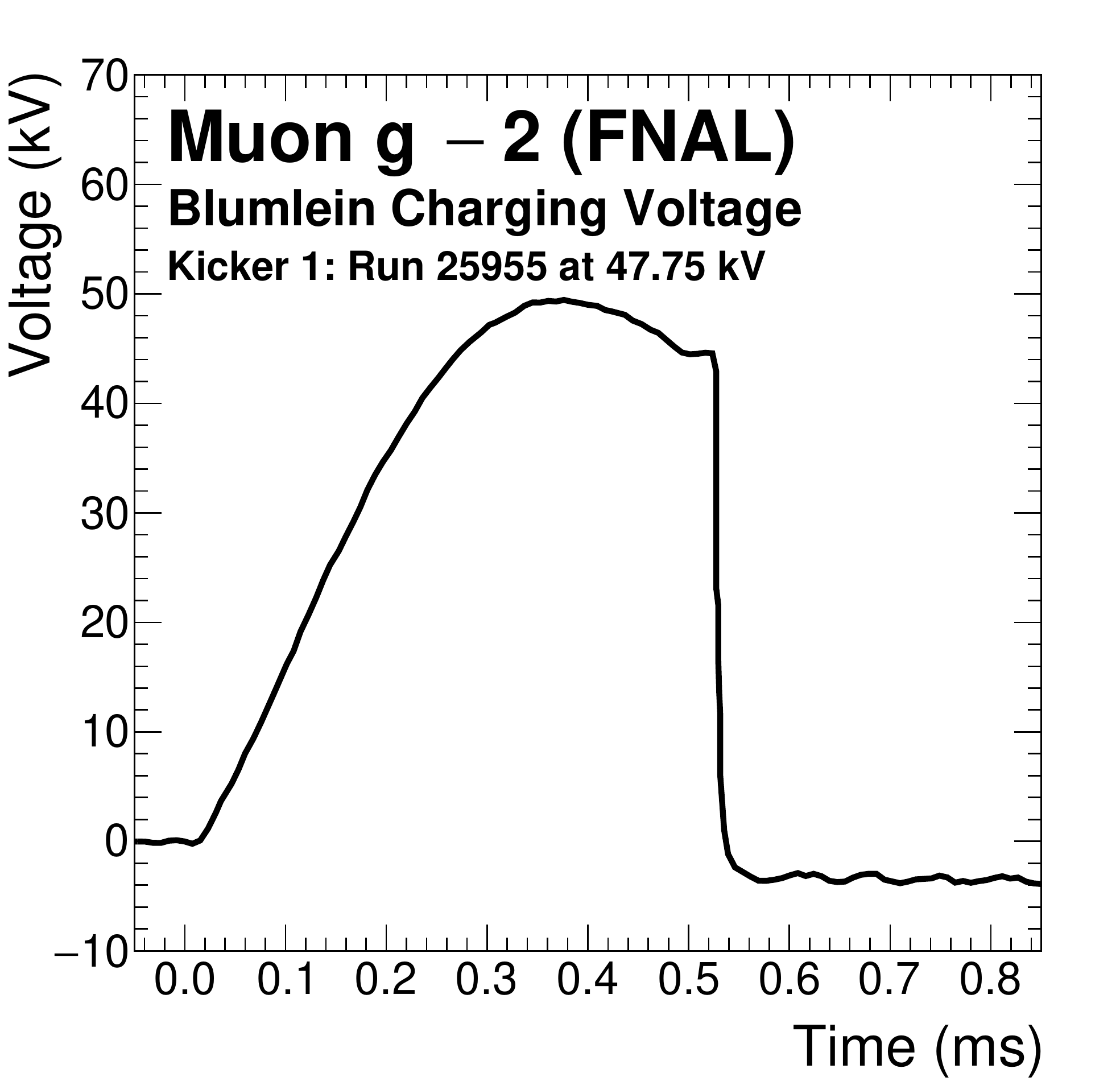}
	\caption{Measurement of the Kicker 1 Blumlein PFN charging voltage taken from the secondary winding of the K1 transformer. The sharp drop in voltage at around t = 0.52 ms comes from the thyratron switching element entering the conductive state. This transition is what generates current in the load. Kicker 1 was operating at 47.75 kV at the time of observation, which can be derived from the magnitude of the voltage drop induced by the thyratron switch. \label{fig:SV}}
\end{figure}

During production running, we obtained essential measurements by monitoring voltages from the PFN charging systems and a trio of pickup coils installed near the SRV leads. Each custom-made, single-loop coil is constructed using copper coaxial line with a Teflon dielectric, and the areas of the loops are approximately 3.14 cm$^2$. From the PFN voltage waveforms, we monitored information regarding the trigger and timing controls, the health of the Blumlein pulsers, and the kick amplitude. An example from the data is shown in Fig.~\ref{fig:SV}, where the PFN voltage is measured through a capacitive divider attached to the secondary winding of the charging transformer. Key information pertaining to the trigger system can be deduced through this waveform~\cite{ipac18}. The SCR mentioned in Subsection 2.1 initiates the rising edge of the pulse, and the thyratron entering a conductive state generates the sharp voltage drop at around t = 0.52 ms. Any absence or time deviation in either of those features indicates a potential issue with the trigger controls. Likewise, variations from the typical shape displayed in Fig.~\ref{fig:SV} hint at potential electrical breakdowns inside the Blumlein PFN.

We also analyzed and calibrated the strength of the kick amplitude by using the magnitude of the voltage drop in the PFN charging voltage data. We present a more rigorous description of the calibration analysis in the following section. 
\begin{figure}[]	
	\includegraphics[width=\columnwidth]{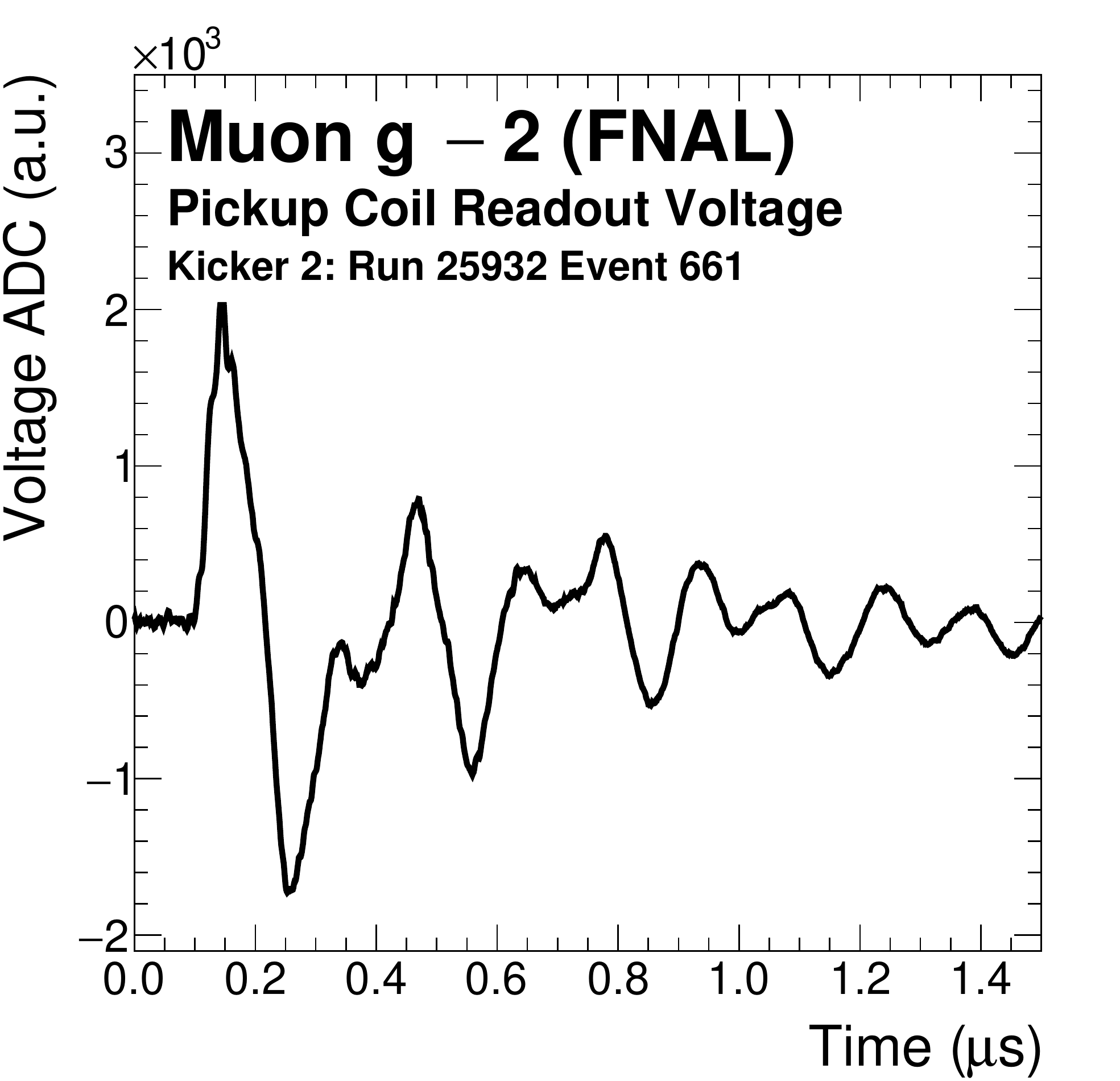}
	\caption{Readout voltage from a pickup coil located near the leads of a kicker magnet. The shape of the waveform depends on the current pulsing through the leads. We use this signal as the primary method to detect sparks, as anomalous currents significantly change the pulse structure. \label{fig:bdot}}
\end{figure}
\begin{figure}[]
	\includegraphics[width=\columnwidth]{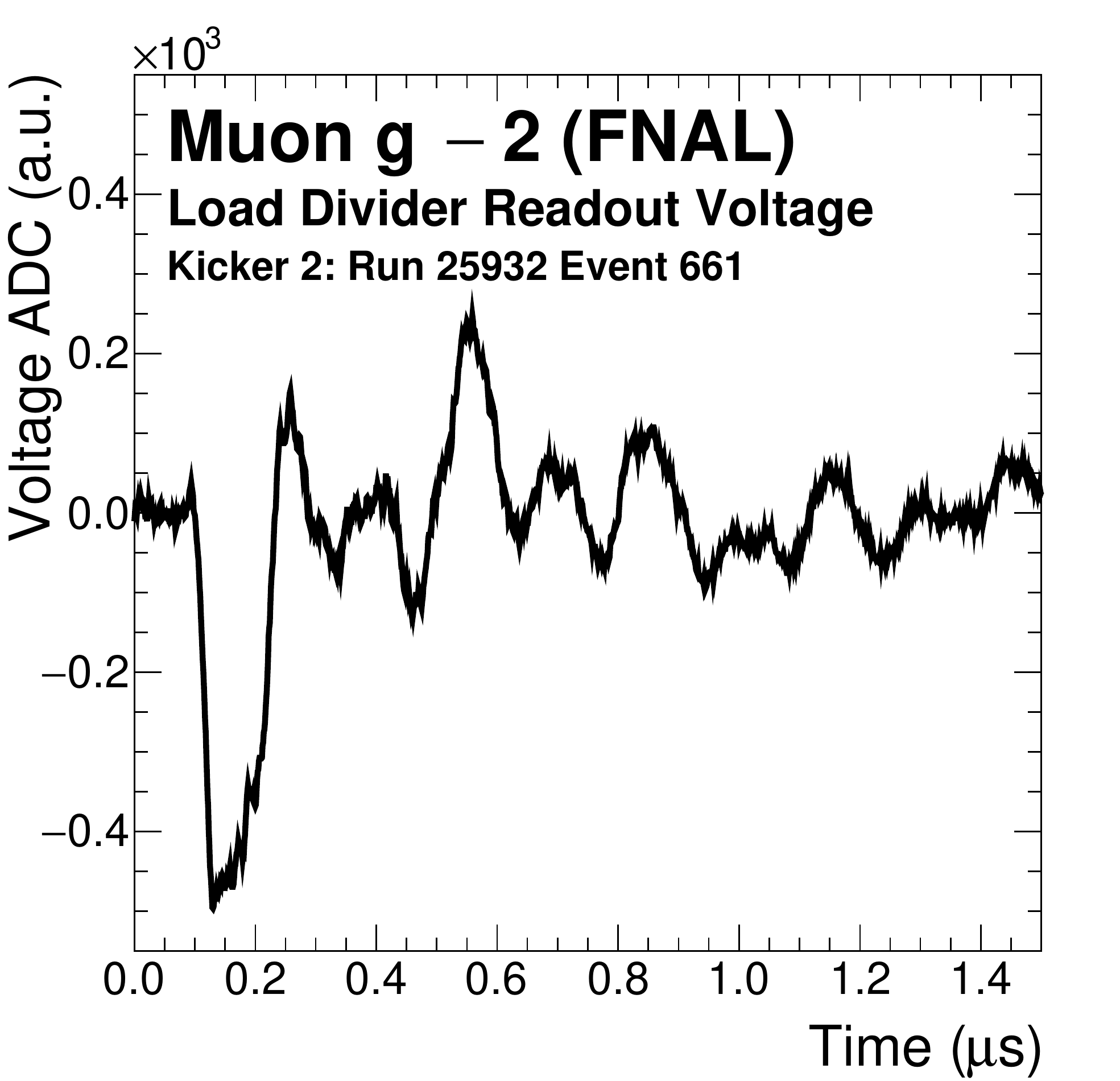}
	\caption{A typical voltage measurement across one of the resistive loads. Data are read out through a capacitive divider with a ratio of roughly 10000:1. The vertical axis units are arbitrary as there is an additional scaling factor generated by the data acquisition hardware.\label{fig:bcd}}
\end{figure}

The readout of the pickup coil offers another important suite of monitoring information. At a fundamental level, the pickup coil data are related to the current on the high voltage leads that connect an SRV feedthrough to the plates of a kicker magnet. Subsequently, the pickup coil facilitates measurements of the electrical pulses that actually perturbate the magnetic field in the beam storage region. While we do not use this monitor to calibrate the magnetic field strength due to miscellaneous noise sources, the integrated pickup coil signal is related to the magnetic field generated by the kicker system.

We utilize the pickup coil signals in the optimization of the kicker timing.  From this waveform, see Fig.~\ref{fig:bdot}, we determine when each kicker produces its maximum field. This event corresponds with the first zero crossing in the pickup coil pulse, which should ideally align with the arrival of the muon beam. With this information, the trigger controls can be adjusted accordingly. Fits performed on data collected during kicker timing scans indicate that a tolerance of $\pm 5$ ns satisfies the experiment's beam storage requirements.

The primary function of the pickup coils is to serve as spark (electrical breakdown) monitors. Any spark in the kicker system impacts the current such that a notable shape change in the waveform occurs. For example, a spark in the SRV will produce a large distortion on the shape shown in Fig.~\ref{fig:bdot}, and corresponding vacuum behavior is usually recorded. A breakdown in the Blumlein PFN will not only distort the PFN charging voltage waveform, but it will also significantly shift the timing of the pulse in the pickup coil data. Consequently, we feed the outputs from the pickup coils into dedicated boards in the controls rack. If the boards register an anomalous shape change in the signal, they automatically veto subsequent charging to prevent damage to the system.

The shape of the pickup coil waveforms, specifically the reflections in the tail structure, generated additional analyses. The inductances of the kicker magnets and HV leads produce the reflections seen in the readouts. This finding poses a significant challenge for two reasons. First, we opted to use the Blumlein PFNs given the timing specifications of the experiment, and the load inductance affects that performance. Second, we required that eddy currents generated by the kicker system become negligible less than 20 $\mu$s after muon injection~\cite{tdr}. The presence of reflections in the current waveform yields the possibility that the criterion is not being met. In the following section, we describe measurements of the load impedances and magnetic field perturbations to quantify the system characteristics and potential effects.

For Run-2, we installed a new monitor as part of the upgrade to the resistive load. Capacitive dividers facilitate direct voltage measurements from the loads during normal operations. From the waveform in Fig.~\ref{fig:bcd}, we can deduce if there are any bad connections in the line, degraded cables, or malfunctioning components through variations from the typical shape. Under normal conditions, we expect to observe a negative voltage pulse that is generated by one of the Blumlein PFNs.

\section{\label{sec:level1}System Characterization}
During the commissioning phase, several fundamental measurements were made to characterize the kicker system. These observations can be partitioned into categories pertaining to {\it calibration}, {\it assembly performance}, {\it field study}, and {\it beam dynamics}.

\subsection{\label{sec:level2}Calibration}

The purpose of the kicker calibration is to relate the references from the slow controls to the voltages on the Blumleins at the time the thyratrons conduct ($t_{thyratron}$). The reference is commonly referred to as the {\it set point voltage} ($V_{ref}$) and has a range of $0-10$ volts. As mentioned previously, the voltage on a Blumlein is measured from a capacitive divider attached to the secondary winding of the associated transformer. However, it was determined that the gains and the behaviors of the capacitive dividers differed from transformer to transformer. 

We temporarily installed a calibrated, resistive divider onto the input cable to the Blumlein PFN. At several values of $V_{ref}$, we recorded the voltage waveforms at the calibrated divider and the transformer's capacitive divider. When the thyratrons conduct, a sharp voltage drop in these waveforms is observed. Fitting the magnitude of $\Delta V$ at $t_{thyratron}$ as a function of $V_{ref}$ yielded linear calibration formulas for all three kickers that are valid between operating voltages of $10-60$ kV. Likewise, we obtained the calibrations of the capacitive dividers through these measurements.

During Run-1, the electronics and software controls underwent multiple iterations that altered the $V_{ref}$ scales and behaviors. In addition, we observed nonlinearity in the kicker response when pulsing at progressively higher voltages, and we determined that the Lambda HV supplies could not uniformly charge the PFN when pulsing at the nominal beam repetition rate. These findings motivated the redesign for Run-2 and generated a different approach for calibrating the kicker system operating voltage during the first production run. Through beam dynamics and muon storage studies, we estimate the total kick strength was between 125 and 137 kV during the Run-1 data sets. 

Equations~\ref{eq:1cal}$-$\ref{eq:3cal} provide the Blumlein charging voltages (in kV units) for the respective kicker magnet. These equations apply only to the system configuration established at the start of Run-2. 
\begin{eqnarray}
V^{II}_{K1}(V_{ref}^{K1}) &= &7.26 [kV/V] \cdot V_{ref}^{K1}-0.59[kV],\hspace{5pt}\label{eq:1cal}\\
V^{II}_{K2}(V_{ref}^{K2}) &= &6.78 [kV/V] \cdot V_{ref}^{K2}-0.38[kV],\label{eq:2cal}\\
V^{II}_{K3}(V_{ref}^{K3}) &= &7.18 [kV/V] \cdot V_{ref}^{K3}-0.81[kV].\label{eq:3cal}
\end{eqnarray}

\subsection{\label{sec:level2}Assembly Performance}

In conjunction with the Run-2 upgrades, we made considerable efforts to quantify the performance of the refurbished and redesigned kicker assemblies. Prior to the construction and installation of the new resistive loads, we connected two of the Blumlein PFNs to a test load and recorded the electrical current responses at several kicker operating voltages.

The test load consisted of four 50 $\Omega$ resistors in parallel and a T\&M Research Products Inc. A-4-0025 current viewing resistor (CVR). To account for any parasitic capacitance and inductance, a Keysight impedance analyzer was used to fit the load response to an RLC circuit model. We selected a simple circuit topology in which the resistor and inductor are connected in series, and the capacitor is connected in parallel. Consequently, we can represent the test load in simulation software with a 12.76 $\Omega$ resistor, a 65 nH inductor, and a $6-25$ pF capacitor. The used CVR model has a 2.5 $m\Omega$ resistance, 8 ns rise time, and 48 MHz bandpass. 

\begin{figure}[!h]
	\includegraphics[width=\columnwidth]{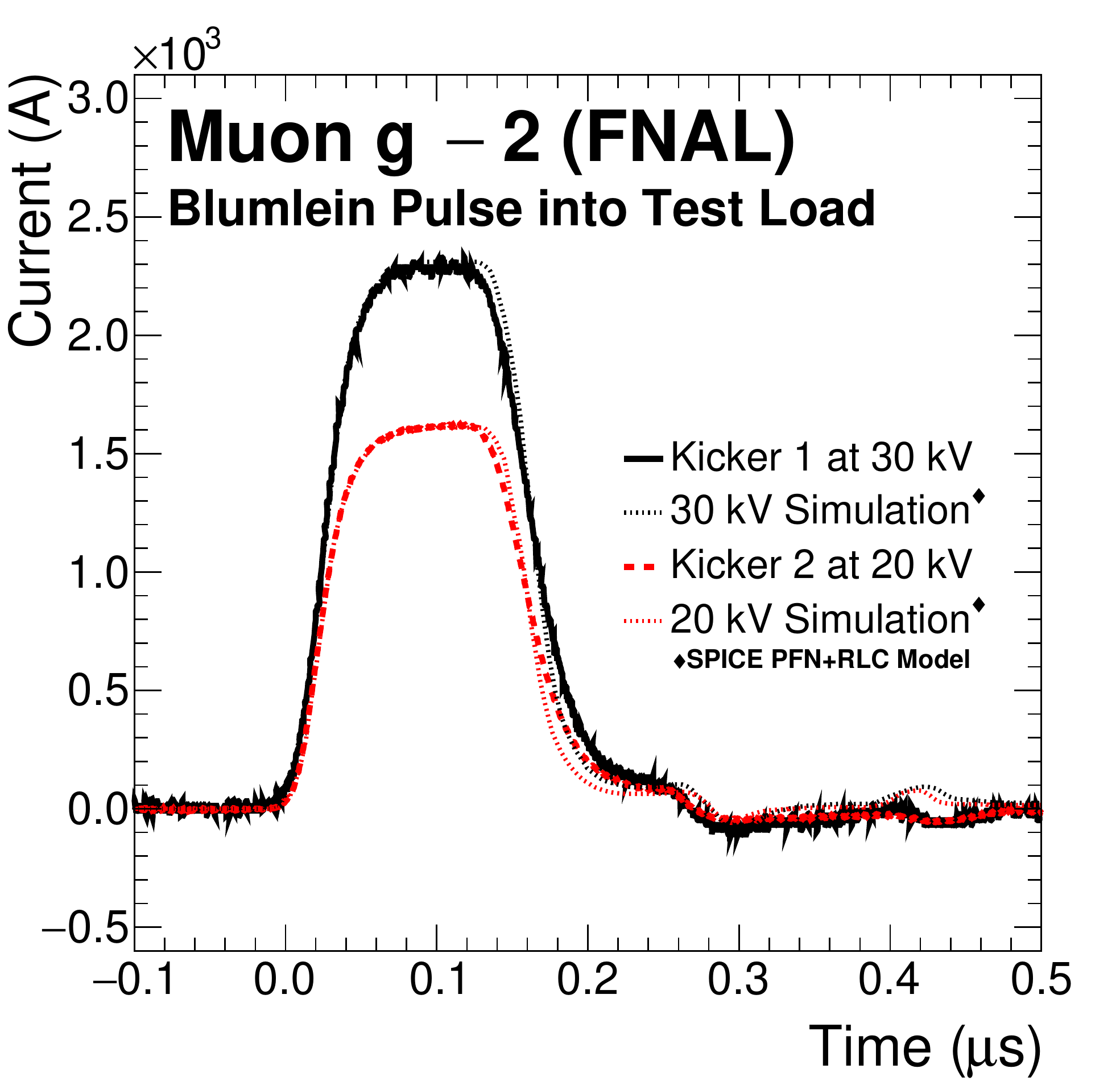}
	\caption{Current through a test load when driven by the Kicker 1 and Kicker 2 Blumlein PFNs. Kicker 1 was operated at approximately 30 kV, and Kicker 2 was operated at approximately 20 kV. SPICE model simulations of the PFN and test load produced the additional waveforms in the figure.  \label{fig:testload}}
\end{figure}

We again utilized SPICE to model the Blumlein PFNs and test load as lump elements. The simulation results are shown in Fig.~\ref{fig:testload} along with the data collected when operating Kickers 1 and 2 at approximately 30 kV and 20 kV, respectively. Space constraints in the experimental hall prohibited measuring the test load response to Kicker 3.

The data agree reasonably well with the results from simulation. Uncertainties in the analyzer RLC fit and the absence of lossy elements in the SPICE model explain the small discrepancies between the data and simulation waveforms. There is an additional uncertainty in the modeling of the thyratrons, and their rise times have been tuned in the SPICE simulation to reflect some minor differences in the thyratron behaviors. Another source of tuning comes from inductive elements in the thyratron portion of the model circuit. We selected thyratron inductances (20 nH for Kicker 1 and 15 nH for Kicker 2) by using the amplitudes of the undershoots in the tails of the data waveforms as constraints. We also used the results shown in Table~\ref{tab:trans} to characterize the transformer attributes in the simulation.

The results presented in Fig.~\ref{fig:testload} established that the refurbished Blumlein PFNs behave as expected, as no irregularities in the pulse shape were found. We built further confidence by considering the pulse durations. The Blumlein PFNs measure roughly 9.35 meters. Recalling Eq.~\ref{eq:time}, we compute an expected pulse duration of 133.3 ns, which is typically ascribed to the full width at half maximum (FWHM) of the waveform. Taking the $\tau_{FWHM}$ measurements from data and simulation, we construct Table~\ref{tab:blumtime}. The idealized, simulated, and observed values agree to better than 5$\%$. Uncertainties in our measurements of the relative permittivity of castor oil, and the presence of stray inductances absent in Eq.~\ref{eq:time}, explain these differences. 

We also note the shape variation between the waveforms in Figs.~\ref{fig:bazcomp} and~\ref{fig:testload}. The broader current pulse observed in Fig.~\ref{fig:bazcomp} is produced by the larger inductance of the kicker magnets compared to that of the test load. Reflections observed in the pickup coil data (see Subsection 2.6) further corroborate this point, which motivated the following characterization analysis.

Using a Keysight analyzer, we measured the impedances of the kicker load circuits. These series circuits consist of the resistive loads, SRV feedthroughs, and kicker magnets. Impedance data were recorded for driving frequencies ranging from 1 kHz $-$ 5 MHz. We selected this frequency domain based on the behavior of the Blumlein PFNs and the pickup coil readouts, from which the rise time of the waveforms provide an upper bound for the relevant frequencies.
\begin{table}[]
	\centering
	\caption{\label{tab:blumtime}Blumlein pulse durations ($\tau$) taken at the FWHM of the current waveforms shown in Fig.~\ref{fig:testload}. Comparison to idealized expectation is also shown.}
	
	\begin{tabular}{llll}
		\hline
		\hline
		Kicker &$\tau_{ideal}$ &$\tau_{simulation}$ &$\tau_{data}$\\
		\hline
		K1 &133.3ns &137.6ns &137.8ns\\
		K2 &133.3ns &136.8ns &139.0ns\\
		\hline
		\hline
	\end{tabular}
	
\end{table}
\begin{figure}[!h]
	\includegraphics[width=\columnwidth]{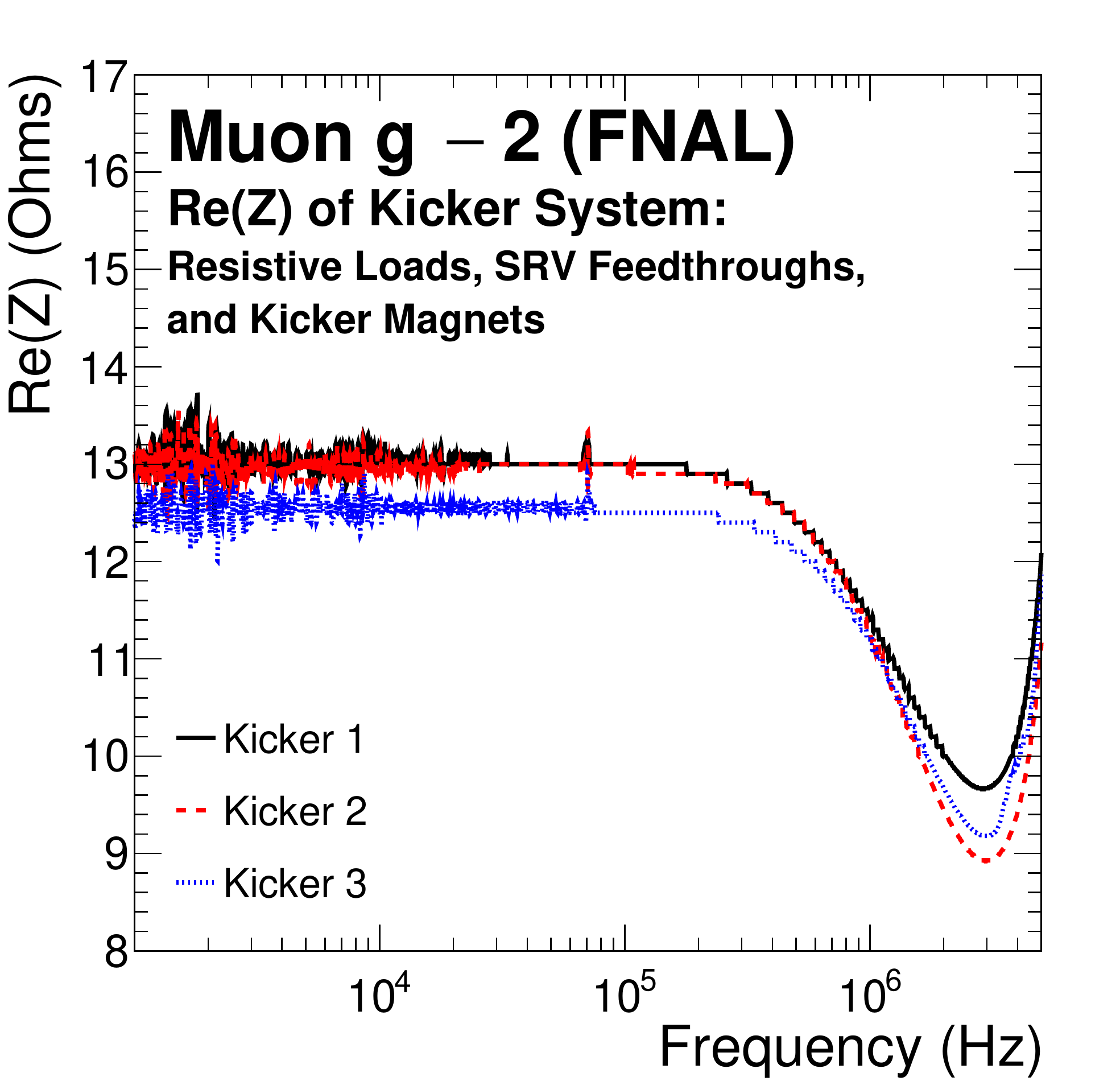}
	\caption{The real impedance component of a series circuit consisting of the resistive load, SRV feedthrough, and kicker magnet. A Keysight impedance analyzer drove the circuit over a 1 kHz $-$ 5 MHz range for these measurements. The response is shown for all three kickers. \label{fig:ReZ}}
	
	\includegraphics[width=\columnwidth]{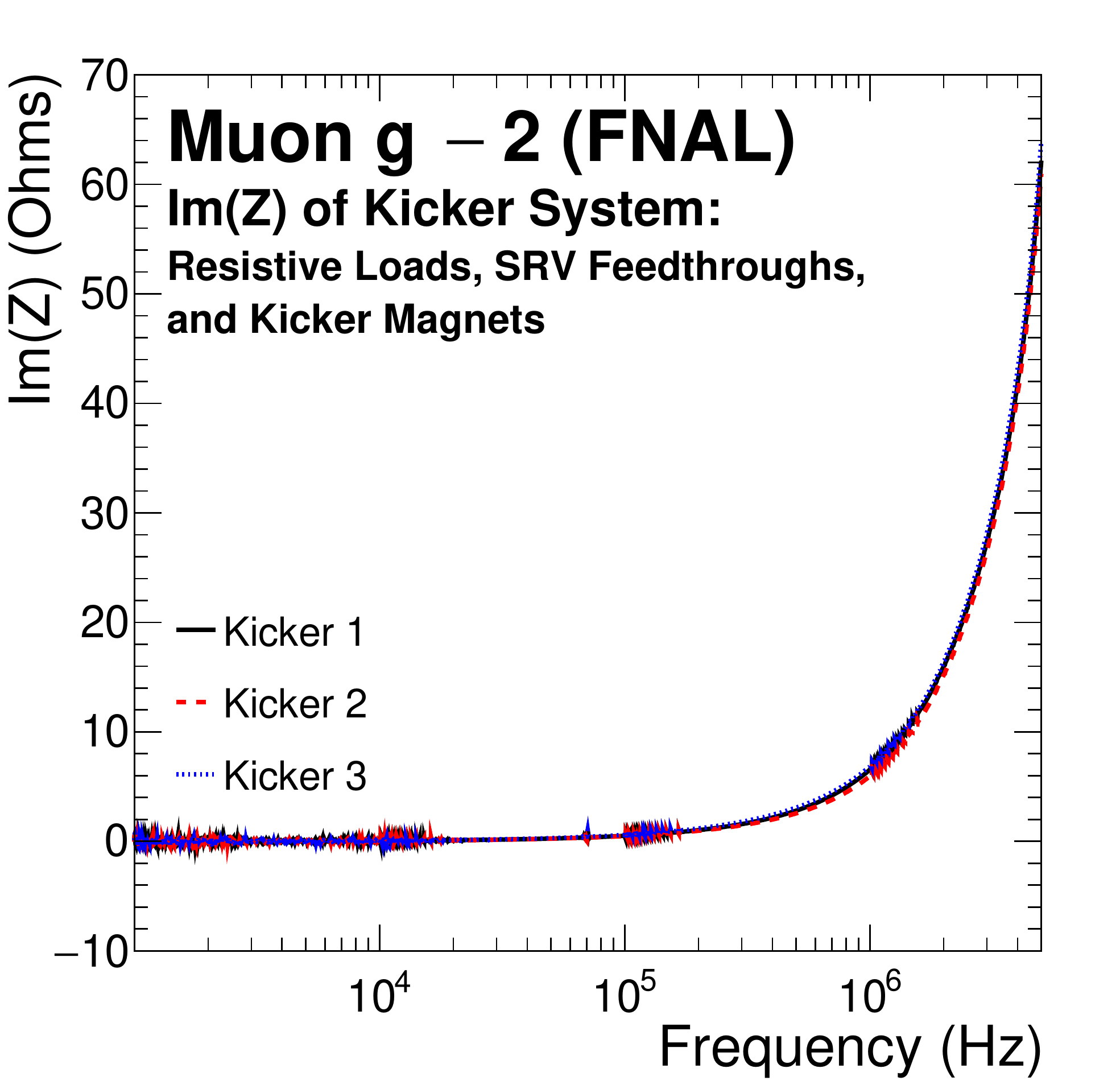}
	\caption{The imaginary impedance component of a series circuit consisting of the resistive load, SRV feedthrough, and kicker magnet. A Keysight impedance analyzer drove the circuit over a 1 kHz $-$ 5 MHz range for these measurements. The response is shown for all three kickers. \label{fig:ImZ}}
\end{figure}

Figure~\ref{fig:ReZ} shows the real component of the impedance as measured in the three kicker loads. First, we found that the impedance converged to 12.5 $\Omega$ $-$ 13.0 $\Omega$ at low frequencies. This behavior fits expectations when considering the topology of the system. At low frequencies, the inductive kicker magnet has a negligible impact on the total impedance, and the capacitive network inside the resistive load will be fully charged at a time $t << T_{analyzer}$, where $T_{analyzer}$ indicates the period of the driver oscillator. In this case, the parallel network of eight 100 $\Omega$ resistors provides all of the load impedance, and the circuit resistance should converge to 12.5 $\Omega$. The measurements in this frequency regime were consistent with the resistor tolerances.

At higher frequencies in the 1 kHz $-$ 5 MHz domain, we observe a minimum in the real component of the impedance that demonstrates the measurable effect of the capacitive network in the Run-2 load topology. The minimum in all of the measurements shown in Fig.~\ref{fig:ReZ} occurs at approximately 3 MHz. We employ rudimentary circuit knowledge to understand the location of these minima. Considering only the RC components of the load, the associated timing constant of 45 ns corresponds to a frequency of 3.5 MHz. Likewise, the resonant frequency of the LC components in the load is around 2.3 MHz. The actual circuit consists of non-lump elements and parallel paths that complicate the picture. However, these two frequencies straddle the minima at similar values of Re(Z) and can be looked at as conservative bounds on the minima positions. Furthermore, we observe $\approx$75-ns rise times for the two Blumlein pulsers in Fig.~\ref{fig:testload}, which corresponds to a similar underlying frequency of 3.$\bar{3}$ MHz. Overall, this specific measurement provides tangible evidence that the goal of the load topology transition discussed in Subsection 2.3 has been realized.

Figure~\ref{fig:ImZ} shows the imaginary component of the kicker load impedance in the 1 kHZ $-$ 5 MHz frequency domain. We observe notable behavior at higher frequencies, where the inductance of the kicker magnet yields the dominant contribution to the total impedance of the load. At 3 MHz, the measured load inductance is between 1.4 $\mu$H and 1.5 $\mu$H. The characterization corroborates several findings from simulation and data. The inductance of the load is responsible for the distortion of the main current pulse when comparing Figs.~\ref{fig:bazcomp} and~\ref{fig:testload}. Furthermore, the generated impedance mismatch with the Blumlein PFN yields reflections that are clearly observed in the pickup coil waveforms, Fig.~\ref{fig:bdot}. The experiment has made considerable efforts to understand the impact of the reflections on the data, particularly under the purview of beam simulation.

\subsection{\label{sec:level2}Field Study}

As mentioned in Subsection 2.6, several monitors record electrical waveforms from which the Muon $g-2$ Experiment can assess the status of the kicker systems. These devices are crucial for operations, but they do not directly measure the magnetic field. We constructed a Faraday magnetometer to evaluate the kicker-induced field amplitude and waveform shape in a similar manner to the procedure described in Section 4 of~\cite{og}.

\begin{figure*}[]
	\centering
\includegraphics[width=1.9\columnwidth]{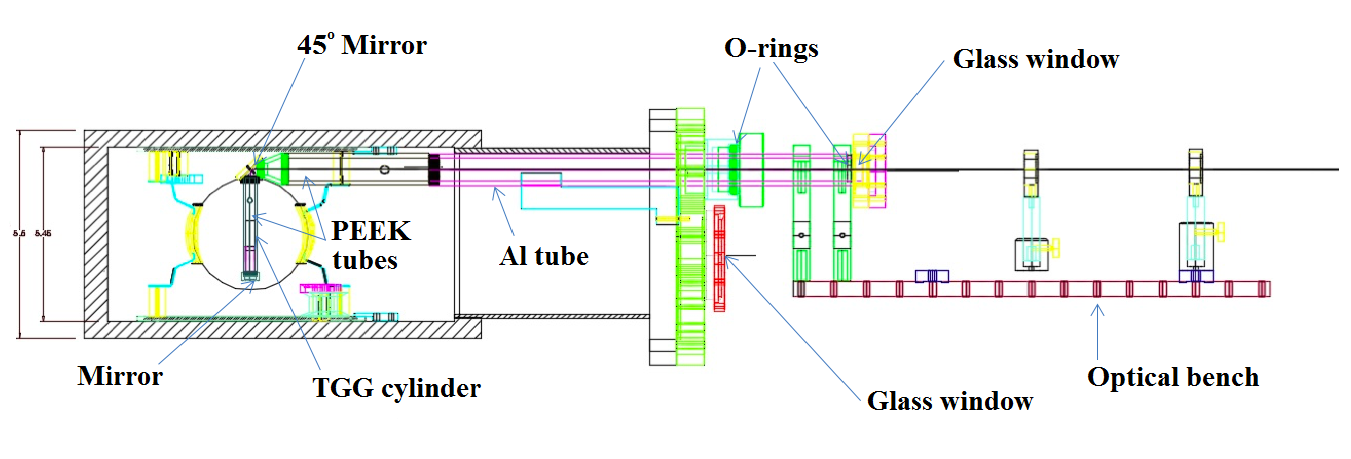}
\caption{Diagram of the magnetometer apparatus once installed in the kicker magnet region. From left to right, the principal features of this diagram include the kicker magnet, the TGG cylindrical housing, the magnetometer arm, the interface with the SRV, and the optical bench.\label{fig:magarm}}
\end{figure*}

The magnetometer exploits the physics of the Faraday Effect and Malus's Law. The former of these principles, described by Eq.~\ref{eq:far}, states that certain materials can change the polarization angle of linearly polarized light in the presence of a parallel magnetic field. In addition to the field strength, the change in polarization angle depends upon the Verdet constant ($v$), which quantifies the strength of the Faraday Effect in the material, and the distance light travels through that medium ($d$).
\begin{equation}
\Delta\theta = vBd. \label{eq:far}
\end{equation}

In this analysis, we used a Terbium gallium garnet (TGG) crystal as the Faraday rotator, which has a Verdet constant of approximately 126 [rad $T^{-1} m^{-1}$] for 639 nm light. Laser light was set incident on a linear polarizer to establish an initial polarization for the beam, which we define here to be $\theta_{0} = 0$. The light then passed into the SRV through the arm of the magnetometer before a 45$^{\circ}$ mirror directed the laser light down into the crystal cylinder, which was placed along the midplane of the kicker region. A second mirror, situated at the bottom of the crystal housing, reflected the laser light back through the crystal and out the magnetometer arm.

Upon exiting the SRV region, the laser light passed through a second polarizer with $\theta_{p} = \pi/4$, where $\theta_{p}$ is relative to $\theta_{0}$. This particular choice for $\theta_{p}$ comes from consideration of Malus's Law, Eq.~\ref{eq:mal}.
\begin{equation}
I = I_{0}\cos^{2}\theta, \label{eq:mal}
\end{equation}
where the intensity of light passing through a polarizer is related to the initial intensity ($I_{0}$) and the angle of polarization ($\theta$). From the derivative, Eq.~\ref{eq:dmal}, we find that angular perturbations around $\theta_{p} = \pi/4$ yield the largest changes in the measured intensity, and thus, this configuration optimizes the signal sensitivity. Further adjustments to $\theta_{p}$ were routinely applied. For example, data taking while the 1.45 T ring field was active necessitated a shift to account for the Faraday rotation generated by the static field. 
\begin{equation}
\frac{dI}{d\theta} = -2I_{0}\cos\theta\sin\theta. \label{eq:dmal}
\end{equation}

\begin{table*}[!h]
	\centering
	\caption{\label{tab:magm}Magnetometer measurements of the midplane magnetic field in comparison to the expectation from simulation. The mean baseline voltage and mean detector voltage response are also included where relevant. Note that subsamples were subject to different gain settings, which explains the change in absolute scale observed between sources.} 
	
	\begin{tabular}{lrrc}
		\hline
		\hline
		Source &$\overline{V}_{B=0}$ (mV) &$\overline{\Delta V}$ (mV) &$\vert B_{midplane}\vert (Gauss)$\\
		\hline
		37-Hour &$767.0\pm43.0$ &$109.3\pm5.3$ &${\bf 217.4\pm16.3}$\\
		Scope &$817.8\pm81.7$ &$121.0\pm3.4$ &${\bf 225.6\pm23.7}$\\
		\hline
		Simulation & $-$ & $-$ & ${\bf 217.1\pm16.6}$\\
		\hline
		\hline
	\end{tabular}
	
\end{table*}

Combining Eqs.~\ref{eq:far} and~\ref{eq:mal}, we arrive at the expression shown in Eq.~\ref{eq:pmal}, where $l$ is effectively the length of the TGG crystal. The expression for the kicker-induced magnetic field is derived and shown in Eq.~\ref{eq:bmal}, where we make use of the $B = 0$ condition to relate the field strength to the laser intensity observed when the kicker magnet is off ($I_{B=0}$) and the maximum change in intensity induced by the magnet pulse ($\Delta I$).
\begin{eqnarray}
I &= &I_{0}\cos^{2}(\pi/4+\Delta\theta),\nonumber\\
 &= &\frac{I_{0}}{2}(1-\sin(2\Delta\theta)),\nonumber\\
 &= &\frac{I_{0}}{2}(1-\sin(2vBd)),\nonumber\\
 &= &\frac{I_{0}}{2}(1-\sin(4vBl)). \label{eq:pmal}
\end{eqnarray}
\begin{eqnarray}
B &= &\frac{1}{4vl}\sin^{-1}(1-2I/I_{0}),\nonumber\\
& &I(B=0) = I_{0}/2,\nonumber\\
\
\therefore B &= &\frac{1}{4vl}\sin^{-1}(-\Delta I/I_{B=0}).\label{eq:bmal}
\end{eqnarray}

We measure the signal through the use of a Thorlabs PDA10A(-EC) Si-amplified fixed gain detector, which receives the laser light after it passes through the second polarizer. The detector produces a voltage waveform that is directly proportional to the input power, as expressed in Eq.~\ref{eq:thor}. 
\begin{equation}
V = R(\lambda)G_{t}\frac{R_{load}}{R_{load}+R_s}P,\label{eq:thor}
\end{equation}
where $R(\lambda)$ is the responsivity of the detector at a given laser wavelength, $G_{t}$ is the transimpedance gain, $R_{load}$ and $R_s$ represent resistances added to the output of the detector, and $P$ is input power~\cite{thor}. The input power is the only dynamic variable in this equation. Thus, we arrive at the final form for the expression of the kicker-induced magnetic field as a function of the detector voltage by exploiting the relationship in Eq.~\ref{eq:thorrel}.
\begin{equation}
\frac{\Delta V}{V_{B=0}} = \frac{\Delta P}{P_{B=0}} = \frac{\Delta I}{I_{B=0}}.\label{eq:thorrel}
\end{equation}
\begin{eqnarray}
\therefore B &= &\frac{1}{4vl}\sin^{-1}(-\Delta V/V_{B=0}).\label{eq:bvmal}
\end{eqnarray}

A diagram of the physical apparatus is included in Fig.~\ref{fig:magarm}. The optical bench upon which the laser, polarizers, and detector were mounted is illustrated on the right side of the image. The arm of the magnetometer, made of aluminum and PEEK ({\it Polyether ether ketone}), passes into the SRV through a special flange. On the left side of the figure, the profile of the kicker magnet is shown along with the cylindrical TGG housing. We used a 1.3-cm-long crystal with a 5 mm diameter for this measurement.

With the magnetometer installed into the SRV and the TGG crystal inserted between the kicker plates, we operated Kicker 1 at 47.75 kV. Additional data were collected during the ramping of the main ring magnetic field, which confirmed the calibration constant of the TGG crystal shown in Eq.~\ref{eq:mcal}.
\begin{equation}
vl \approx (1.64 \times 10^{-4}  \pm 1.36 \times 10^{-6}) \frac{rad}{G}.\label{eq:mcal} 
\end{equation}

Three data sets were collected for the magnetometer studies presented here that include the crystal calibration, a small sample read out with a Tektronix oscilloscope, and a 37-hour sample acquired through the primary DAQ.

\begin{figure}[!h]
	\includegraphics[width=\columnwidth]{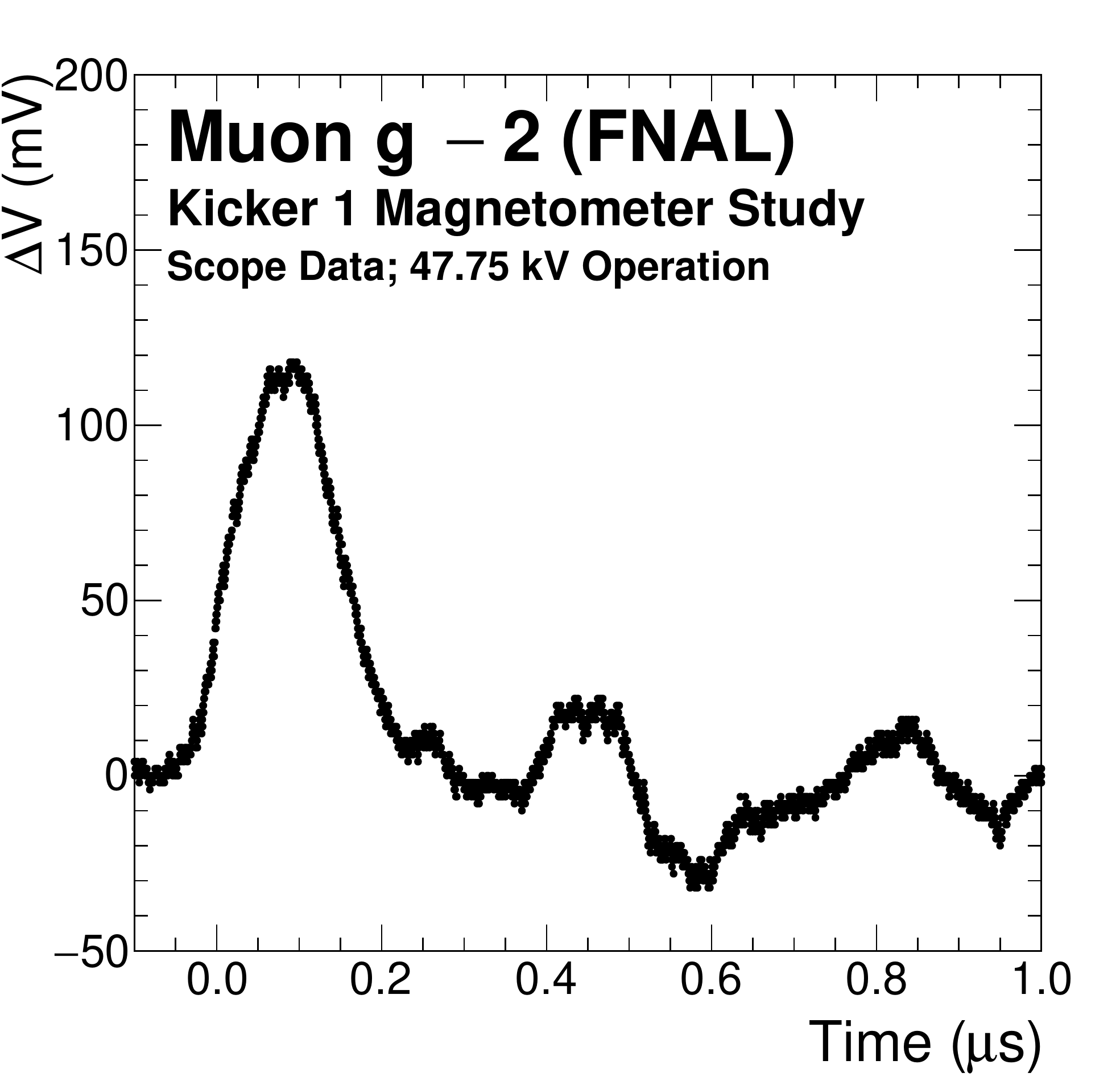}
	\caption{Example voltage waveform produced by the detector of the magnetometer. We extract the peak values of $\Delta V$ over many of these traces to acquire the $\overline{\Delta V}$ provided in Table~\ref{tab:magm}. \label{fig:magp}}
\end{figure}

The detector readout generated the voltage waveform in Fig.~\ref{fig:magp}, which displays $\Delta V(t)$ over the duration of the kicker pulse. $V_{B=0}$ was obtained by measuring the DC-coupled voltage produced by the laser prior to kicker magnet operation. Combining the circuit and field models, we predict that a $217.1\pm16.6$ G midplane magnetic field will be produced by a Blumlein charging voltage of 47.75 kV.  Table~\ref{tab:magm} contains a summary of the study, from which we conclude that the kicker-induced magnetic field produced by a 47.75 kV charge was $217.4\pm16.3$ G.

\begin{figure*}[]
	\centering
	\includegraphics[width=1.9\columnwidth]{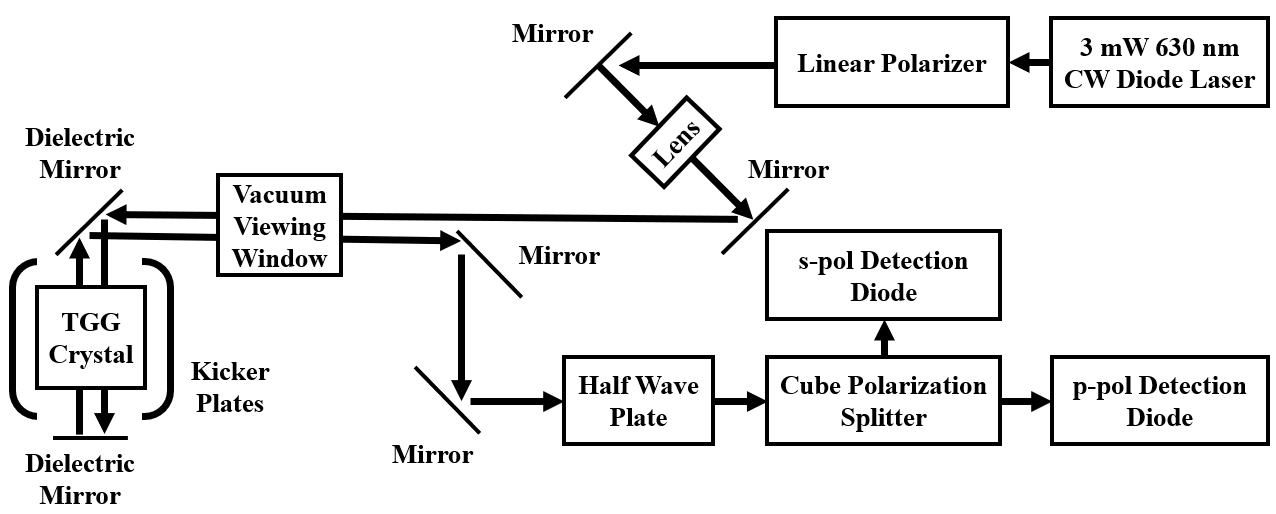}
	\caption{Flow diagram of the 2PD magnetometer assembly used to assess magnetic field transients induced by kicker operation.\label{fig:2PDA}}
\end{figure*}
\begin{figure*}[]
	\centering
	\includegraphics[width=\columnwidth]{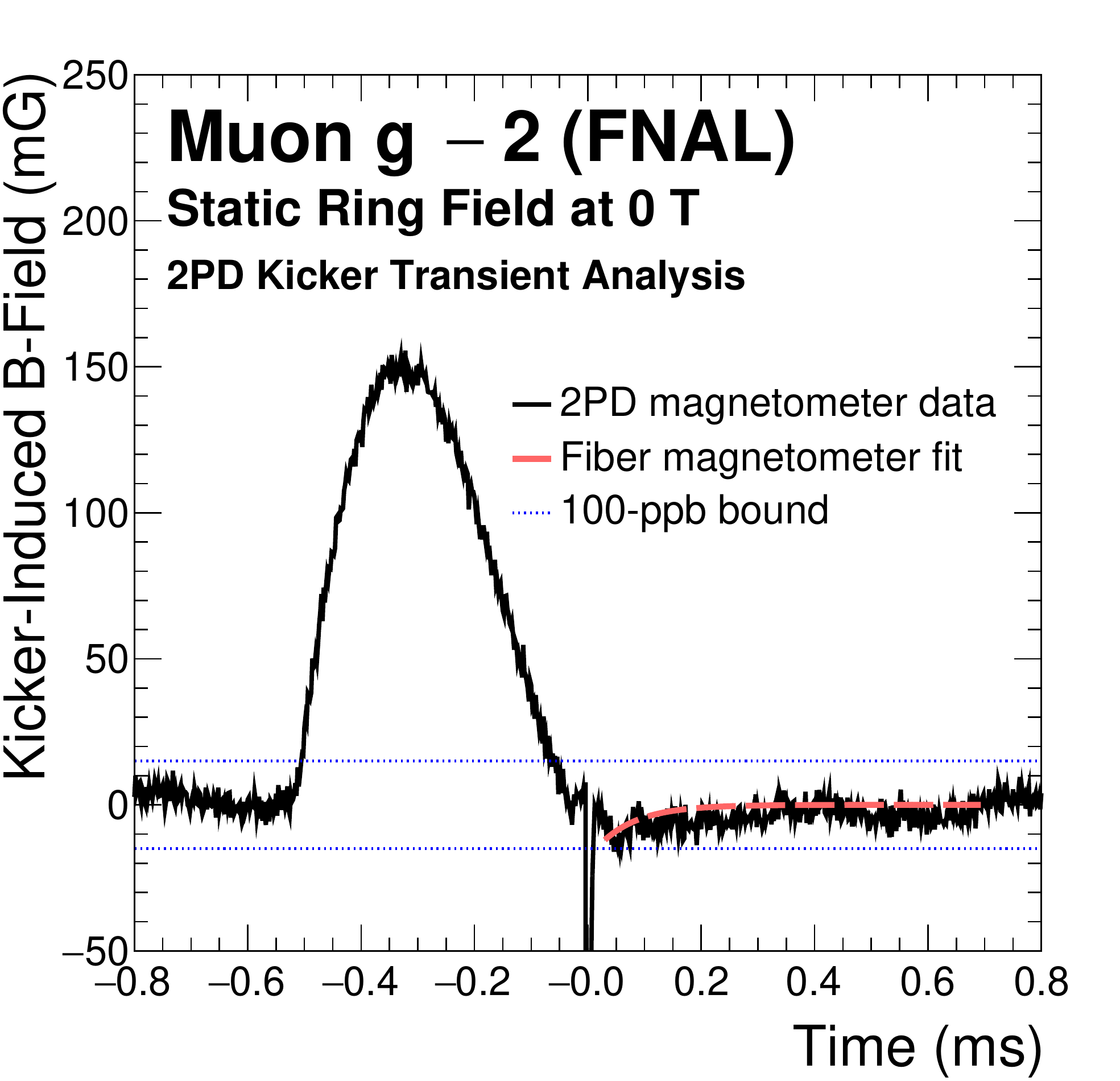}
	\includegraphics[width=\columnwidth]{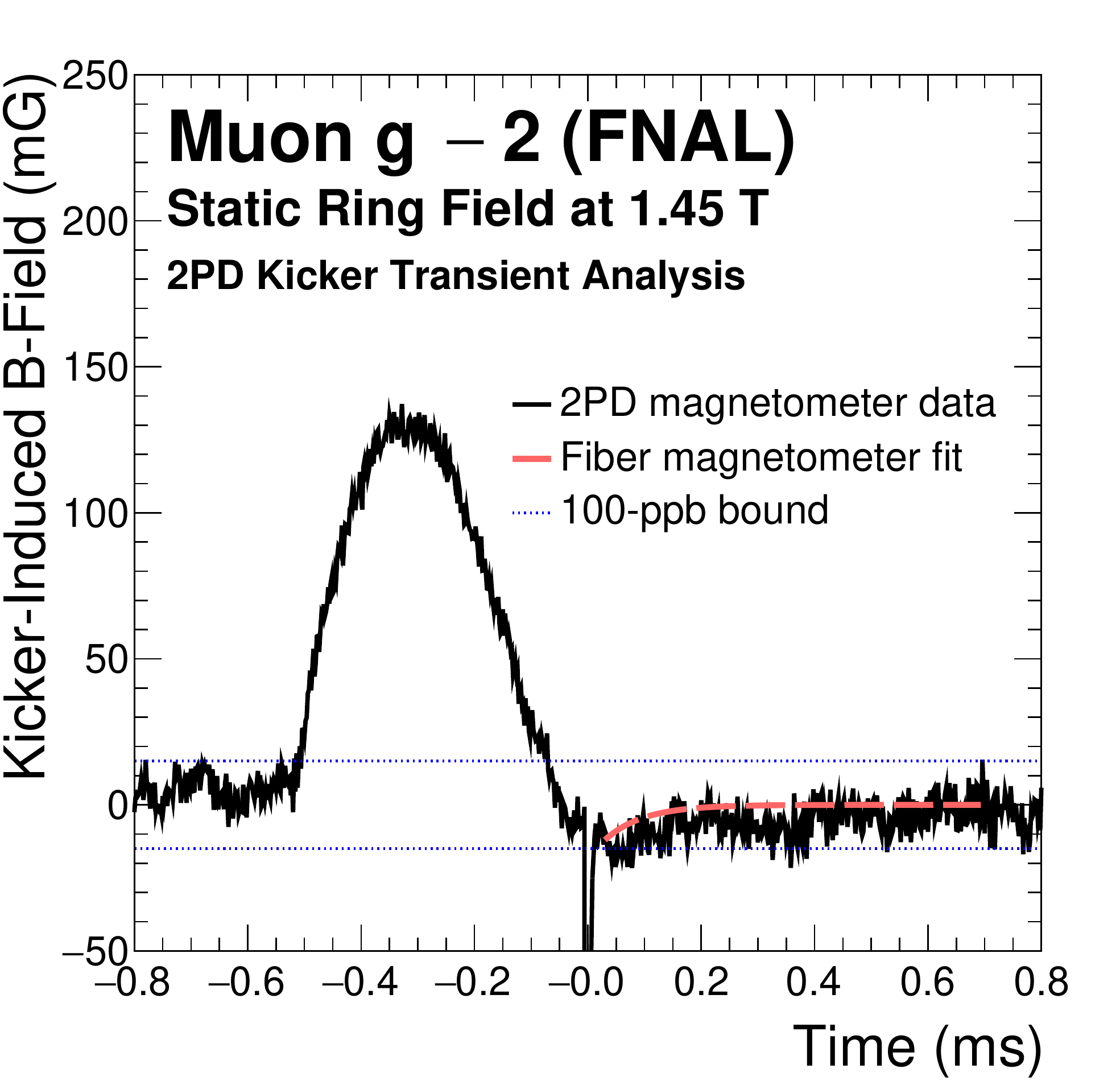}
	\caption{Kicker-induced magnetic field in mG (solid black) vs. time as determined by the 2PD assembly with the static ring field at 0 T ({\it left}) and 1.45 T ({\it right}). We include the fit result from the fiber magnetometer analysis (dashed red) in the time domain of the muon precession measurement. The dotted blue lines correspond to a hypothetical 100-ppb shift on the measured value of $\omega_{a}$ produced by a transient magnetic field.\label{fig:hogan}}
\end{figure*}

Additional magnetometer analyses were conducted to assess the impact of potential, unwanted kicker-induced magnetic fields at time scales much larger than the kicker pulse duration. Both induced eddy currents and line reflections could yield fields capable of perturbing the measurement of $a_{\mu}$. First, a second photodiode and a polarizing beam-splitter were added to the configuration described in this section to produce the 2PD assembly. Through the acquisition of two orthogonally polarized data sets, subtraction methods increased the sensitivity of the measurement in the time domain $t >> 149$ ns. However, we discovered that ambient sources introduced mechanical vibrations in the magnetometer arm that hindered efforts to reach the desired level of precision.

We also deployed a compact magnetometer apparatus that made use of optical fibers, inline beam-splitters, and two TGG crystals to improve the sensitivity to small fluctuations in the magnetic field. A full description of this magnetometer assembly, its calibration and underlying sensitivities (to both magnetic field sources and backgrounds), and the subsequent analysis are provided in Section VIII of~\cite{field}. That study concluded that: 
\begin{equation}
	\frac{\Delta\omega_{a}}{\omega_{a}} = [-27 \pm 37]\space ppb, 
\end{equation}
 where the expression gives the fractional pull on the anomalous precession frequency ($\omega_{a}$) driven by the non-optimal behavior of the kicker pulse. Similar to the BNL E821 result in~\cite{og}, the magnitude of this systematic fits well within the FNAL experiment's long-term precision goal of 140 ppb.
 
 A supplementary result to the fiber magnetometer study emerged following improvements made to the 2PD assembly. Two Thorlabs PD36A2 diodes, with a 15 kV/A transimpedance gain and a 1 MHz bandwidth, were installed onto the aluminum optical bench and terminated into high-impedance waveform digitizers. The flow diagram provided in Fig.~\ref{fig:2PDA} illustrates the setup of the 2PD magnetometer assembly. To further assess and verify the impact of mechanical vibrations on the system, we measured the lateral position and amplitude of the incident light in a dedicated systematic test using an additional set of Thorlabs PDP90A2 diodes.
 
 By adjusting the input laser light polarization, half wave plate, and laser optics, we optimized the 2PD assembly for operation in two static ring field conditions: 0 T and 1.45 T. The sensitivity of the 2PD assembly is roughly 46.9 mV/mT, which can be compared to the more sensitive 68.3 mV/mT level achieved by the fiber magnetometer. Figure~\ref{fig:hogan} shows the kicker-induced field measurements for each of the selected ring conditions. The dotted blue lines at $\pm$15 mG correspond to a hypothetical $\pm$100-ppb shift on the measured value of $\omega_{a}$, and the solid black lines show the measurements of the kicker-induced fields. The primary kicker pulse drives the narrow, negative spikes at t = 0 ms. The Blumlein charging current produces the positive $130-160$ mG perturbations prior to t = 0, which have no impact on the assessment of $a_{\mu}$ as they occur before the muon measurement period. We observed considerably more activity in the tail structure when the ring field operated at 1.45 T, which stemmed from Lorentz forces producing mechanical vibrations in the kicker plates.  
 
 Within the 0.03$-$0.70-ms time window critical for the analysis of muon precession, we measured the average field:
 \begin{equation}
 	B_{2PD} = [-5.1\pm0.3_{stat}\pm0.5_{cal}\pm7.8_{vib}]\space mG, 
 \end{equation}
where the additional terms represent uncertainties due to statistics, calibration, and system vibration. We assigned the vibration term by examining the RMS of the signal $0.5-1.2$ ms before the kicker pulse.  The fit result from the fiber magnetometer analysis (dashed red lines) is also displayed in Fig.~\ref{fig:hogan}. We found agreement between the two approaches, further indicating that transient fields induced by the kicker system did not have a detrimental impact on the Muon $g-2$ measurement.
 
\subsection{\label{sec:level2}Beam Dynamics}

We conclude the system characterization with a brief study of beam dynamics. As a critical ring system, the kicker magnets have a significant impact on the physics data. In particular, there exist relationships between the kicker performance, muon storage, and beam-related systematics that warrant discussion.  

From a statistics standpoint, we note the correlation between the amplitude of the field perturbation (the kick strength) and the number of muons that are successfully stored in the SRV. The significant redesign of the non-ferric kickers between E821 and E989 was motivated by several factors. In addition to the timing characteristics of the Blumlein PFNs, muon storage optimization served as an engineering driver.

Following the expressions in Eq.~\ref{eq:int}~\cite{tdr}, the integrated field perturbation ($Bdl$) is directly proportional to the desired displacement angle ($\beta$). Thus, an optimal perturbation exists for the geometry of the experiment's beam injection apparatus. As shown in Fig.~\ref{fig:geometry}, $\beta\approx 10.8$ mrad for both E821 and E989. The produced value of $Bdl$ in the SRV is dependent upon the charging voltage of the PFN, meaning that there exists an ideal operating voltage about which the muon storage is maximized. Electrical breakdowns in the E821 circuit prevented the experiment from reaching the storage maximum during operation~\cite{tdr}. Consequently, the E989 kicker was designed to provide better storage efficiency.

\begin{eqnarray}
\frac{d\vec{p}}{dt} &= &q\vec{v}\times\vec{B},\nonumber\\
d\vec{p} &= &qvB\hat{r}dt,\nonumber\\
p\beta\hat{r} &= &qBdl\hat{r},\nonumber\\
\therefore Bdl &= &\frac{p\beta}{q}.\label{eq:int} 
\end{eqnarray}

We include two strength scans to quantify the muon storage at various kicker operating voltages. The first data collection occurred at the start of Run-2 once other ring systems were installed and updates to the data acquisition software were functional. The second collection period occurred during Run-3 after the 4x50 $\Omega$ AA5966 cable configuration was replaced with a 3x32 $\Omega$ Dielectric Sciences 2264 cable arrangement.  
\begin{figure}[]
	\includegraphics[width=\columnwidth]{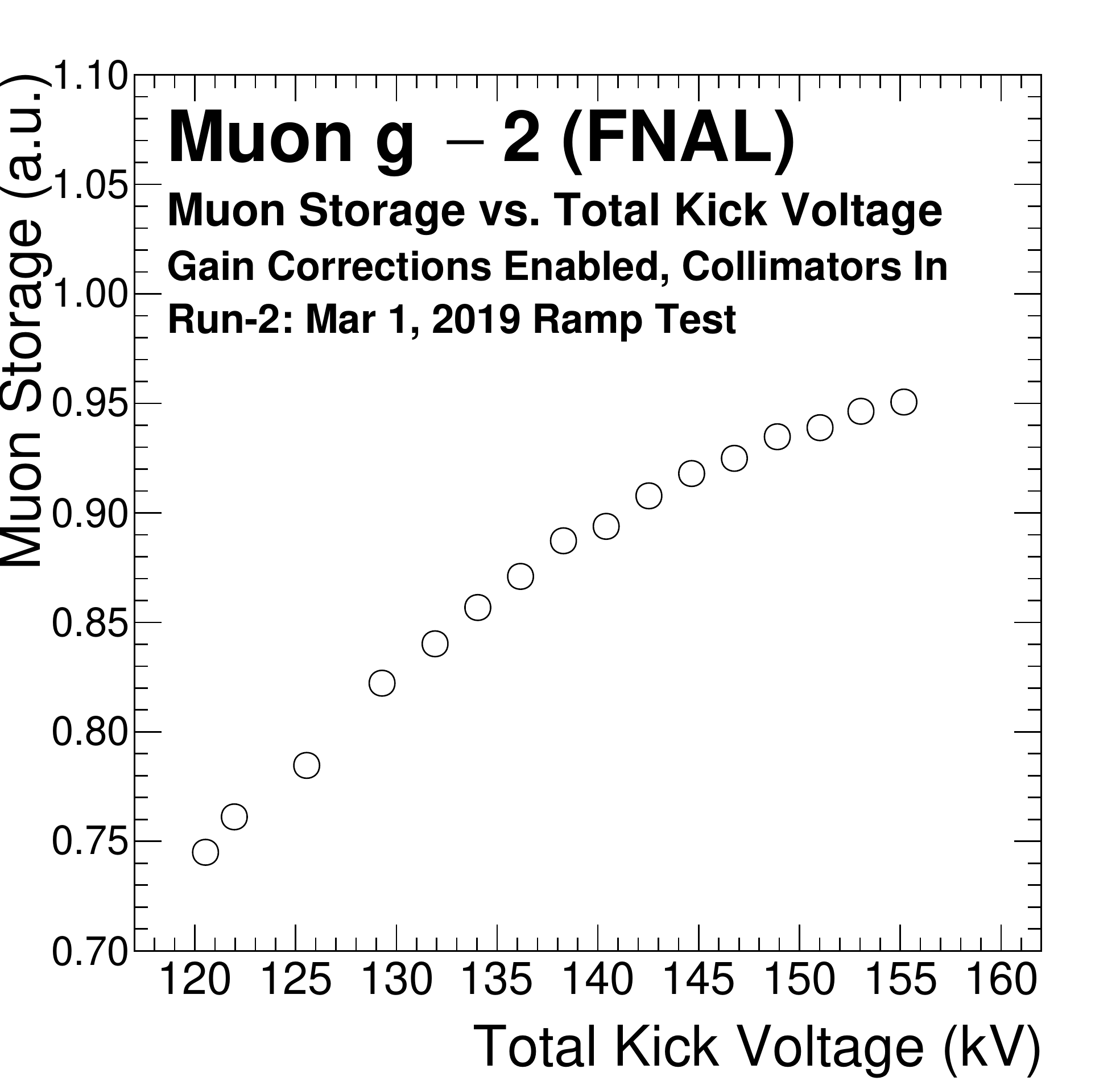}
	\caption{Muon storage, presented in arbitrary units, as a function of the total kick strength in kV. We performed this preliminary scan early in Run-2 as part of the conditioning phase.\label{fig:ms}}
\end{figure}
\begin{figure}[]
	\includegraphics[width=\columnwidth]{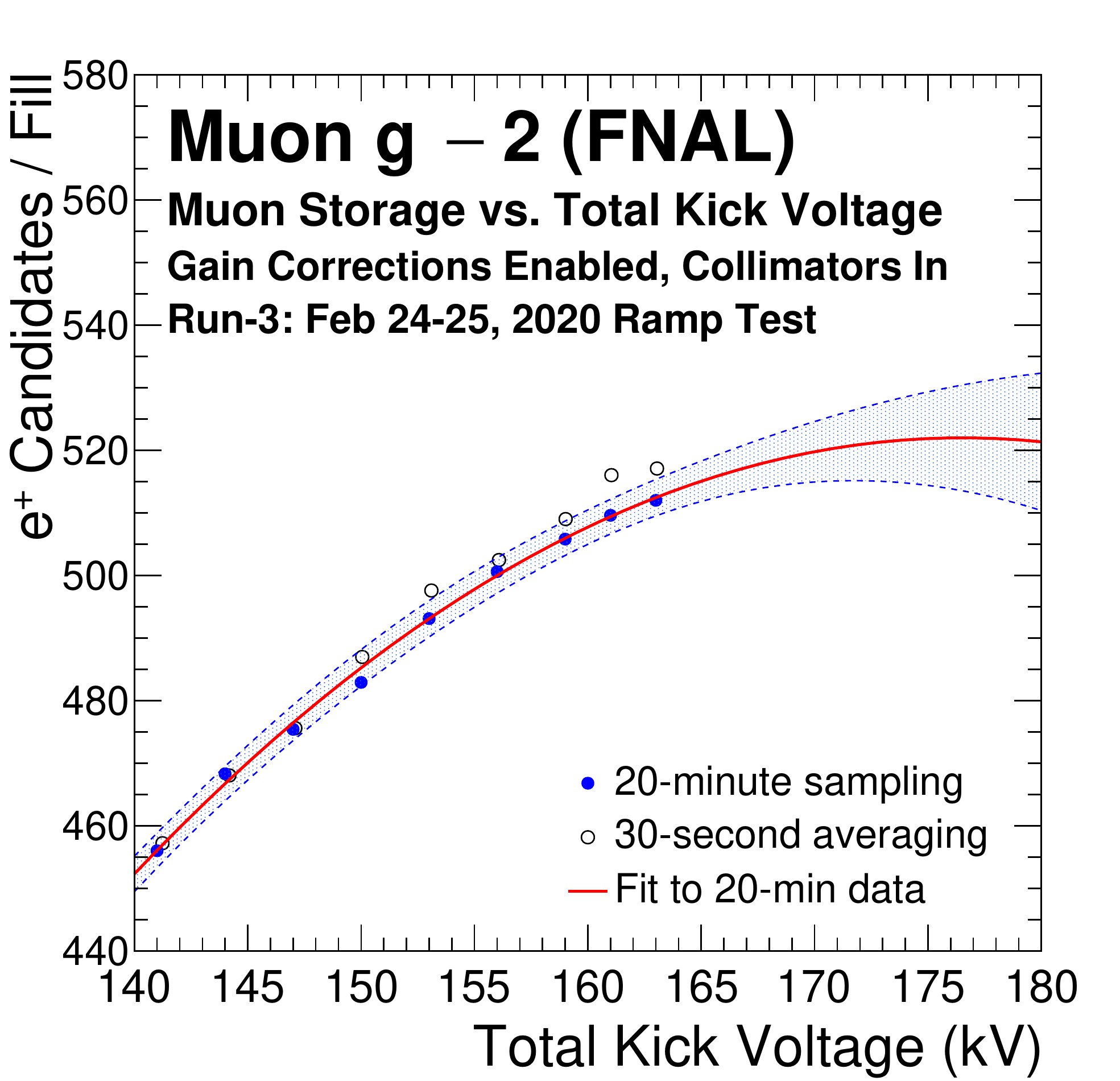}
	\caption{Muon storage, in $e^{+}$ candidates/fill units, as a function of the total kick strength in kV. We performed this scan during Run-3. The blue points represent data collected over 20-minute-long runs at each voltage setting. The red fit line and associated band show the muon storage predicted by this data set. The black circles display the observed muon storage when we average the storage metric over multiple 30-second-long samples. The quadrupoles operated at 18.2 kV during this test, and the inflector magnet current was 2763 A.\label{fig:ms2}}
	
	\includegraphics[width=\columnwidth]{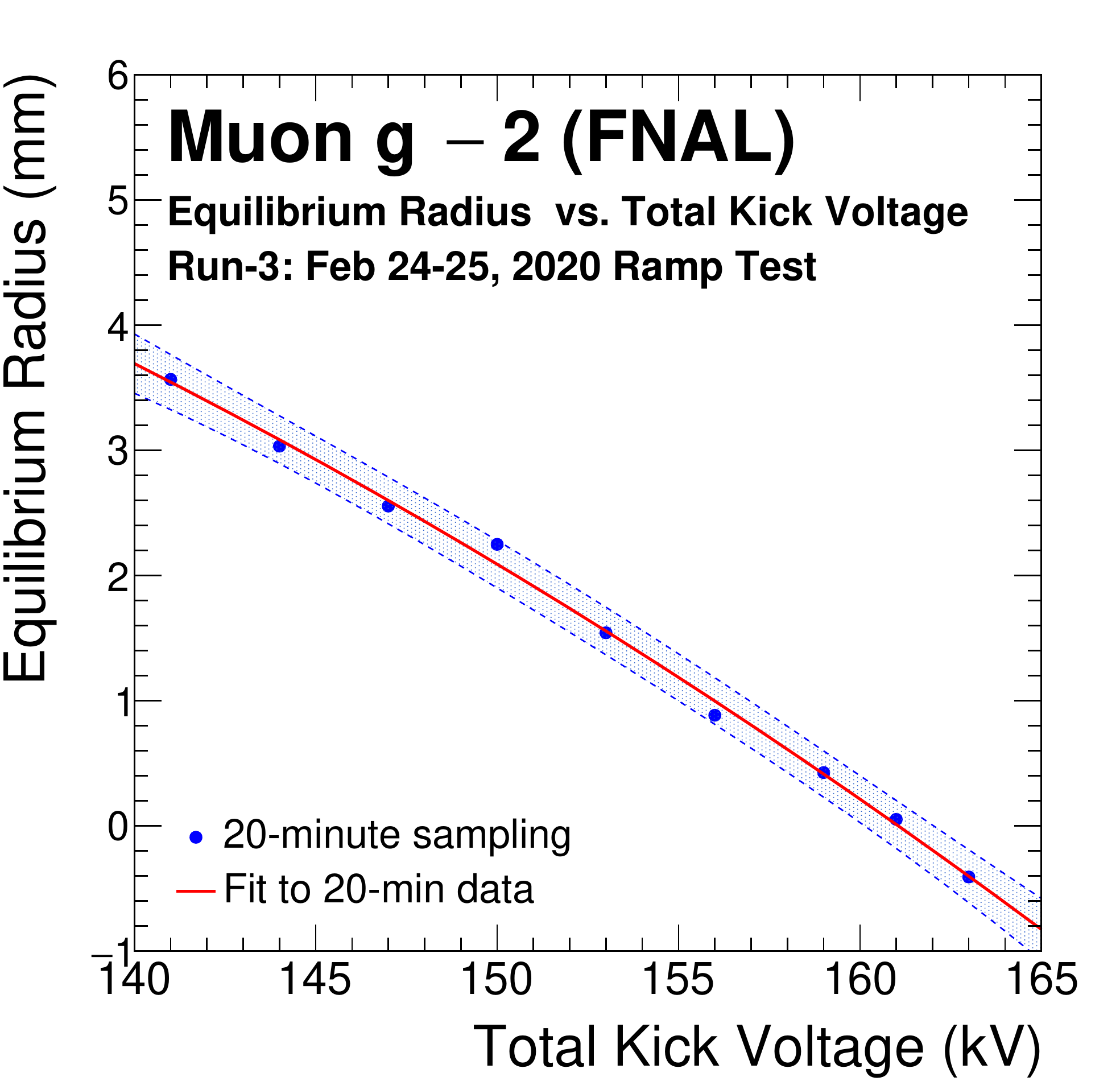}
	\caption{The mean equilibrium radii of the stored muon beam as a function of the total kick strength. The data are represented by the blue points. A fit to the data (red line) and associated uncertainty band are also included in the figure. We find that the beam is centered by a total kick amplitude of approximately 161 kV.\label{fig:xe}}
\end{figure}

The results of the first study, displayed in Fig.~\ref{fig:ms}, yielded the expected behavior given the range of the search. The preliminary assessment in the design report placed the optimal voltage at approximately 165 kV, at which point we expected to observe a maximum in the muon storage distribution. The voltage of interest, represented as the total kick voltage, is the sum of the three individual kicker voltages, and it is proportional to the total induced magnetic field observed by the muons. Starting at 120 kV, we saw an increase in the muon storage as the voltage was raised. At around 140 kV, a change in the rate of rise was noted, suggesting that we were approaching the optimal set point. The test was stopped at 155 kV due to concerns for equipment safety. Despite the limitation, this result represented the first evidence from both E821 and E989 that the muon storage turnover point was being approached.

Figure~\ref{fig:ms2} displays the results of the Run-3 strength scan. Through Run-2, and up until this test, the total kick strength was fixed to 142 kV to reduce the breakdown frequency of the AA5966 cables. Following the installation of the 2264 cables, we were able to improve operating conditions by studying the newly accessible portions of the parameter space. The muon storage distributions were analyzed using two methods, both of which are displayed in the figure. In the first method, we averaged the number of positrons per fill over multiple 30-second intervals at each kicker set point. This approach sacrificed overall statistics, but it facilitated on-the-fly data quality assessments that mitigated the impact of potential beamline variability. The second technique used Gaussian fits over the full data set to determine the positrons per fill observed in 20-minute-long runs at each kicker voltage setting. Both methods yield comparable results and shift the predicted muon storage maximum from 165 kV to $175-182$ kV. We can explain this shift by considering the expression in Eq.~\ref{eq:int}, the actual measurements of the kicker-induced magnetic field shown in Subsection 3.3, and the corresponding simulations. $175-182$ kV charging amplitudes will generate angular displacements of $9.7 < \beta < 10.4$ mrad. We found these calculations to be consistent with the 10.8 mrad specification when considering additional elements that were not included in the design report, such as beam injection and the resistive load topology.

From a systematics standpoint, kicker performance primarily affects the radial distribution of the stored muon beam and coherent betatron oscillations (CBO). Investigation of the radial distribution in Run-1 served as the main evidence that the kicker system had been operating at a lower voltage than expected. The finding was the genesis for many of the upgrade efforts described in this manuscript. The culmination of this work came in Run-3 with the installation of the 2264 cables, which also facilitated a thorough mapping of the radial muon distributions and CBO amplitudes at set points ranging from the Run-2 operating point up to the centering position.

The equilibrium radii at various kicker set points were determined using a Cosine Transform Technique as described in Ref.~\cite{orlov}. These data (blue points) are shown in Fig.~\ref{fig:xe} along with a fit and shaded uncertainty band. We find that the optimal kicker strength for centering the stored muon beam is approximately 161 kV ({\it when the electrostatic quadrupoles operate at 18.2 kV and the inflector magnet current is 2763 A}). The 161 kV set point became the primary kicker operating mode as a consequence of this study.

The measurement of $a_{\mu}$ is affected by electric fields in the muon storage region~\cite{bds} that modify the observed precession frequency. By centering the beam, we reduce the impact of this E-field effect by producing a more ideal muon momentum distribution. Furthermore, the impact of the CBO is greatly reduced by shifting to 161 kV operations. Over the course of the Run-3 scan, we observed a decrease in the CBO amplitude from around 10 mm to 4 mm. We expect this shift will improve the quality of the physics data. For Run-1, the experiment performed rigorous studies to determine the impact of the kicker system performance on the measurement of the muon magnetic anomaly. In Subsection 3.3, we described the effect on the magnetic field measurement detailed in Ref.~\cite{field}. Sections II and III of Ref.~\cite{bds} provide more information pertaining to the calculation of the systematic uncertainties driven by beam dynamics considerations. 

In summary, we have characterized all elements of the Muon $g-2$ Experiment's non-ferric kicker and measured the system impact on several beam dynamics properties. While we have yet to observe the muon storage maximum with respect to the kicker operating voltage, we achieved the ideal mean equilibrium radius during the Run-3 test and confirmed that the upgrade efforts led to improved conditions for our analyses.

\section{Conclusions}
The fast non-ferric kicker system developed for the Muon $g-2$ Experiment facilitated the efficient capture of muons for the production of multiple large data sets. Compared to BNL E821, these new data collectively contain roughly nine times the number of decay positron events, from which we will produce the most precise measurement of the muon magnetic anomaly to date. During these operations, we discovered several limitations in the kicker system that motivated upgrades between Run-1 and Run-2. System characterization studies presented in this manuscript demonstrated the effectiveness of those improvements along with general reliability. We observed that the non-ferric kickers behaved within expectation, and we have determined that remaining limitations do not significantly impact the quality of the physics result.

Any future experiment making use of similar technology should revisit the nature of load inductance. The characteristics of the magnet load produced an impedance mismatch in the line that generated unwanted reflections. While this proved to not be detrimental to the physics goals of the experiment, significant resources were invested to quantify the effects and develop potential solution strategies. Invariably, these considerations constructed a complex parameter space spanning the number of kicker magnets, the inductance of each load, potential circuit capacitance, the pulse width, necessary infrastructure, and cost.

\section*{Acknowledgments}
We thank our Muon $g-2$ collaborators and Fermi National Accelerator Laboratory (Fermilab), a U.S. Department of Energy, Office of Science, HEP User Facility for the resources provided. Fermilab is managed by the Fermi Research Alliance (FRA), acting under Contract No. DE-AC02-07CH11359. The authors are supported by the Department of Energy HEP and NP offices (USA), National Science Foundation (USA), Horizon 2020 research and innovation programme (EU) under the Marie Skłodowska-Curie grant agreements No. 690835 MUSE and No. 734303 NEWS, Istituto Nazionale di Fisica Nucleare (Italy), IBS-R017-D1 of the Republic of Korea, UK NSF, and UK Science and Technology Facilities Council under separate grants.


\end{document}